\PassOptionsToPackage{table}{xcolor} 

\documentclass[aoas,preprint]{imsart}

\RequirePackage{amsthm,amsmath,amsfonts,amssymb}
\RequirePackage[authoryear]{natbib}
\RequirePackage[colorlinks,citecolor=blue,urlcolor=blue]{hyperref}

\usepackage[ruled]{algorithm2e}
\usepackage{tikz}
\usepackage{amsfonts}
\usepackage[table]{xcolor}
\usepackage[mathscr]{eucal}

\startlocaldefs
\endlocaldefs

\begin{document}

\begin{frontmatter}

\title{Sparse matrix linear models \\ for structured high-throughput data}
\runtitle{Sparse matrix linear models for high-throughput data}

\begin{aug}
\author{\fnms{Jane W.} \snm{Liang}
\ead[label=e1]{jwliang@harvard.edu}}\thanksref{t1,m1}
\thankstext{t1}{This work was started when JWL was a summer intern at
UCSF, and continued when she was a scientific programmer at UTHSC.  We
thank both UCSF and UTHSC for funding, and supportive environments for
this work.  We thank Jon {\AA}gren, Thomas E. Juenger, and Tracey
J. Woodruff for granting permission to use their data for analysis.
SS was partly supported by NIH grants GM123489, GM070683, DA044223,
AI121144, and ES022841.}
\address{
Department of Biostatistics \\
Harvard T.H. Chan School of Public Health \\
655 Huntington Ave, Boston, MA 02115-6028 \\
\printead{e1}}
\affiliation{Harvard T.H. Chan School of Public Health \thanksmark{m1}}

\and

\author{\fnms{\'{S}aunak} \snm{Sen} \thanksref{m2,m3}
\ead[label=e2]{sen@uthsc.edu}}
\address{
Department of Preventive Medicine  \\
University of Tennessee Health Science Center \\
66 N. Pauline St, Memphis, TN 38163-2181 \\
\printead{e2}}
\affiliation{University of Tennessee Health Science Center \thanksmark{m2}}
\affiliation{University of California, San Francisco \thanksmark{m3}}

\runauthor{J.W. Liang and. \'{S}. Sen}
\end{aug}

\begin{abstract}
Recent technological advancements have led to the rapid generation of
high-throughput biological data, which can be used to address novel
scientific questions in broad areas of research.  These data can be
thought of as a large matrix with covariates annotating both rows and
columns of this matrix.  Matrix linear models provide a convenient way
for modeling such data.  In many situations, sparse estimation of
these models is desired.  We present fast, general methods for fitting
sparse matrix linear models to structured high-throughput data. We
induce model sparsity using an L$_1$ penalty and consider the case
when the response matrix and the covariate matrices are large. Due to
data size, standard methods for estimation of these penalized
regression models fail if the problem is converted to the
corresponding univariate regression scenario. By leveraging matrix
properties in the structure of our model, we develop several fast
estimation algorithms (coordinate descent, FISTA, and ADMM) and
discuss their trade-offs. We evaluate our method's performance on
simulated data, \textit{E. coli} chemical genetic screening data, and
two \textit{Arabidopsis} genetic datasets with multivariate
responses. Our algorithms have been implemented in the Julia
programming language and are available
at \href{https://github.com/senresearch/MatrixLMnet.jl}{\tt{https://github.com/senresearch/MatrixLMnet.jl}}.
\end{abstract}



\end{frontmatter}

\section{Introduction}

The rise of high-throughput technology has been a major boon for
answering complex biological questions.  Advances in automation,
multiplexing and miniaturization now enable us to perform biological
assays in bulk at vastly lower cost compared to a couple of decades ago.
Examples of such technologies include cDNA microarrays,
next-generation sequencing technologies, and mass spectrometry.  The
rise of these technologies have influenced statistical methods by
posing new questions.  They have also spawned the need for faster
computation, since the size of the data forces the analyst to make
trade-offs between statistical efficiency (or perfection) and
computational feasibility.  A well-known example is the wave of
statistical innovation on multiple comparisons that followed the
adoption of microarrays.

In this note, we consider the problem of modeling structured
high-throughput data as the response variable.  This is the goal of a
wide variety of studies, such as chemical genetic screens using mutant
libraries; eQTL experiments (measuring genome-wide gene expression and
genotype in a segregating population); and metabolomics studies
(measuring a large number of metabolites or chemicals using mass
spectrometry).  The data from these studies can be presented as a
large matrix, with annotations characterizing each row and each column
of this matrix.  For example, in a chemical genetic screen where a
large number of mutant strains are phenotyped in a large number of
conditions, the data can be arrayed with each row representing an
experimental run, and each column representing a mutant; we have
information regarding the environment of each run (row annotations)
and the gene mutated (column annotations).  The row/column annotations
define {\em a priori} known structure in the data.  The goal is to
identify gene-environment interactions in the screen (connections
between row and column annotations), with the underlying idea that
such interactions are rare (row-column connections are sparse).  We
propose to accomplish this using sparse matrix linear models that
provide considerable flexibility in modeling the data. We achieve
sparsity by using an L$_1$ penalty on the model parameters and can
handle situations where the covariate matrices are large.  This model
has broad applicability to a wide range of high-throughput data, and
it has attractive computational properties.

Our approach is to provide a unified sparse linear model framework for
analyzing matrix-valued data where we have row and column
covariates.  This approach generalizes the current approach to such
data, where a two-step procedure is usually followed.  In a microarray
study with two conditions (treatment vs control), the data is in a
matrix, with each row being a sample (row covariates indicate the
sample condition) and each column being a gene.  Genes may be grouped
into pathways (column covariates).  The standard approach is to detect
differential expression contrasting two conditions using t-tests for
each individual gene expression
measurement \citep{dudoit2002microarray}.  These methods have been
extended to situations where each sample may have covariates; instead
of performing a t-test for each gene expression measurement, a linear
model is fit with the covariates as
predictors \citep{ritchie2015limma}.  To understand patterns across
genes or gene groups, a second analysis across genes is performed; for
example gene set enrichment analysis might be performed
\citep{subramanian2005gsea}.  By unifying the two steps into a single
linear model, the analyst gains flexibility in modeling (especially in
the second step where the analysis can have non-categorical or
non-overlapping covariates) and computational speed, as well as power to
detect associations \citep{liang2019matrix}.  In this note, we consider
estimation of these models with a sparsity constraint.

In the Section \ref{sec:stats_frame}, we outline the statistical
framework underlying our model.  We follow by describing some example
datasets to which our methods can be applied and which motivated this
work (Section \ref{sec:data}).  The computational strategy is detailed
in Section \ref{sec:comp_methods}, followed by a section on simulation
studies and analysis of example datasets (Section \ref{sec:results}).
We close by summarizing our conclusions and outlining implications for
future work in Section \ref{sec:discussion}.

\section{Statistical framework}
\label{sec:stats_frame}

Suppose that $Y_{n \times m}$ is a response matrix, with rows
annotated by covariate matrix $X_{n \times p}$ and columns annotated
by covariate matrix $Z_{m \times q}$.  Consider the linear model
\begin{equation}\label{eq:mlm}
Y = XBZ'+E, 
\end{equation}
which is equivalent to 
\begin{equation}
    y_{ij} = \sum_{k=1}^p \sum_{l=1}^q x_{ik} \; z_{jl} \; b_{kl} +
    e_{ij} \label{eq:sumeqn}, 
\end{equation}
where the entries of a matrix $A$ are denoted by $a_{ij}$. The matrix
$B_{p \times q}$ contains the coefficients that need to be estimated,
and $E_{n \times m}$ are the errors (Figure \ref{fig:diagram}).  For
example, in a high-throughput chemical screen of a library of mutants,
the response matrix $Y$ would consist of the colony sizes from growing
the library of mutants in a variety of chemical conditions.  Each row
would be a separate run of the experiment; each column would represent
a specific genetic mutant strain.  The matrix $X$ would consist of
information on the nature and doses of the chemical media in which the
mutants were grown, and the matrix $Z$ would have information on which
gene was mutated.  The linear model allows us to model the effect of
both the genes and the chemicals on colony size.  The model is an
example of a bilinear model.  If $Z$ is an identity matrix, the model
reduces to fitting a linear model with the same row covariates ($X$)
for all columns of $Y$.  Likewise, if $X$ is an identity matrix, the
model reduces to fitting a linear model to each row of $Y$ using the
column covariates ($Z$).  Models of this form are also found in
functional regression if the response function is observed on a
regular grid \citep{ramsay2005functional}.

\begin{figure}[ht]
\begin{center}
\begin{tikzpicture}
\filldraw [fill=white, draw=black] (-3,-4) rectangle (0,0); 
\node at (-1.5,-2) {\Large $X$};
\node at (-1.5,-2) {\Large $X$};
\filldraw [fill=white, draw=black] (0.25,-4) rectangle (6,0); 
\node at (3.05,-2) {\Large $Y$};
\node at (3.05,-2) {\Large $Y$};
\filldraw [fill=white, draw=black] (0.25,0.25) rectangle (6,3); 
\node at (3.05,1.5) {\Large $Z'$};
\filldraw [fill=white, draw=black] (-3,0.25) rectangle (0,3); 
\node at (-1.5,1.5) {\Large $B'$};

\node at (-1.5,-4.4) {$p$};
\node at (3.05,-4.4) {$m$};
\node at (6.3,1.5) {$q$};
\node at (6.3,-2) {$n$};
\end{tikzpicture}
\end{center}
\caption{{A visualization of the response ($Y{:}n{\times}m$),
row covariate ($X{:}n{\times}p$), column covariate ($Z{:}m{\times}q$)
and coefficient ($B{:}p{\times}q$) matrices for a matrix linear
model.} The dimensions shown are for illustration only and not
necessarily to scale. $X$, $Y$, and $Z$ are taken from the data, and
the goal is to estimate $B$.}
\label{fig:diagram}
\end{figure}

Note that our model may be expressed in its vectorized form as
follows.  If $\hbox{vec}$ is the vectorization operator that stacks
columns of a matrix into a single column vector, we can write equation
(\ref{eq:mlm}) as
\begin{equation}\label{eq:vecmlm}
\text{vec}(Y) = (Z\otimes X) \cdot \text{vec}(B) + \text{vec}(E). 
\end{equation}
This is in the form of the familiar linear regression model $y
= \mathbf{X}
\beta + \epsilon$, where $y = \text{vec}(Y)$, $\mathbf{X} = Z\otimes X$,
$\beta = \text{vec}(B)$, and $\epsilon = \text{vec}(E)$.  While the
two representations are mathematically equivalent, representing the
data as a matrix linear model using Equation (\ref{eq:mlm}) is
computationally much more efficient.  Even if the covariate matrices
$X$ and $Z$ are moderately large, their Kronecker product can be
prohibitively large.  We elaborate more on this point later in this
section.

Each row of the design matrix $X$ corresponds to a row in the response
matrix $Y$, analogous to how each row of the design matrix
$\mathbf{X}$ corresponds to an observation in the response vector $y$
for a univariate problem. The matrix $X$ can contain an intercept
(column of 1s), as well as any number of continuous or categorical
covariates, such as contrasts encoding for different
conditions. Similarly, the rows of the design matrix $Z$ correspond to
the columns of $Y$, and $Z$ can be specified with an intercept as well
as other covariates with associated contrasts.  It is useful to
consider the consequences of the model specification via the summation
representation of the model expressed in Equation (\ref{eq:sumeqn}).
The estimated coefficient corresponding to the row and column
intercept is the overall intercept in the vectorized model.  The
coefficients corresponding to the intercept of $Z$ and (non-intercept)
columns of $X$ are row main effects, or the regression coefficients if
we were performing univariate regression of each column of $Y$ on the
variables in $X$.  Similarly, coefficients corresponding to the
intercept of $X$ and (non-intercept) columns of $Z$ are column main
effects, or the regression coefficients if we were performing
univariate regression of each row of $Y$ on the variables in $Z$.  The
coefficients corresponding to non-intercept columns of $X$ and $Z$ can
be interpreted as interaction terms between row and columns covariates
--- this is clearer when looking at Equation (\ref{eq:sumeqn}).

We consider the scenario when the entries in $E$ are independently
distributed with mean zero and the same variance.  The estimation
process reduces to finding the least squares estimates, which have a
closed-form solution that can be computed quickly even with a
high-dimensional $X$ or $Z$ matrix \citep{liang2019matrix,
xiong2011flexible}.
If the rows are independent and identically distributed, but the
columns are correlated, then we can transform the data so that the
entries are uncorrelated.  In this scenario,
the covariance structure is of the form
$$V(\hbox{vec}(E)) = \Sigma \otimes I,$$ where $\Sigma$ is the
covariance matrix of the columns.  If $\Sigma$
is known, we can multiply the response by the inverse of the square
root of this matrix, to reduce it to the uncorrelated form assumed
by our model.  If the covariance matrix is unknown, we can replace
$\Sigma$ by an estimate (e.g. from the residuals of a least squares
fit) to perform the decorrelation.

In many problems, $B$ is expected to be
sparse, or we may want to use a sparse $B$ for prediction and
interpretation.  In such settings, sparsity can be induced by adding
the convex LASSO or L$_1$ penalty $\lambda ||B||_1$ to the least
squares criterion \citep{tibshirani1996regression}:

\begin{equation}
  \frac{1}{2} ||Y - XBZ'||_2^2 + \lambda ||B||_1, 
\end{equation}

\noindent for which no closed-form solution exists and standard 
unconstrained optimization methods cannot be applied
\citep{schmidt2009optimization}. Several approaches
for solving the univariate problem are well-established.  When the
covariate matrix $\mathbf{X}$ is high-dimensional (as in genome-wide
association studies), \citet{wu2008coordinate} and
\citet{friedman2010regularization} proposed cyclic coordinate descent. 

Proximal algorithms are another general approach for convex
optimization in non-smooth, constrained, large-scale, or distributed
problems. They use the proximal operator of a function to solve convex
optimization sub-problems in closed form or with standard methods. The
proximal operator of a closed proper convex function $f$ scaled by
$\rho$, sometimes called the proximal operator of $f$ with parameter
$\rho$, is defined as
\begin{equation}
    \text{prox}_{\rho f}(u) = \arg\min_x \left[ f(x) + \frac{1}{2\rho} || x - u ||^2 \right]. 
\end{equation}
$\rho$ controls the speed at which $\text{prox}_{\rho f}(u)$ moves
points toward the minimum of $f$ relative to staying close to $u$. It
is advantageous to use a proximal algorithm when the proximal operator
is simple but the original function is complex.

An example of such proximal algorithms is fast iterative shrinkage and thresholding algorithms
(FISTAs) \citep{beck2009fast}, which come from computer science
literature and have been applied to univariate L$_1$-penalized
regression.  More generally, FISTA is a proximal gradient method
proposed for the nonsmooth, convex optimization problem for a
parameter vector $\theta$ given by a split objective function
\begin{equation}
\min_{\theta}\{h(\theta) = f(\theta) + g(\theta)\}, 
\end{equation}

\noindent where the loss function $f$ is a smooth convex function that
is continuously differentiable with a Lipschitz continuous gradient, 
and the penalty term $g$ is a continuous convex function that may be
nonsmooth and has a simple proximal operator. The alternating
direction method of multipliers (ADMM), also known as Douglas-Rachford
splitting, is another type of proximal algorithm that utilizes the
same objective function split. ADMM is efficient when both of the
separate proximal operators for $f$ and $g$ are easy to
evaluate \citep{parikh2014proximal}. We have implemented extensions of
these three approaches for our multivariate setup, in which $f(B)
= \frac{1}{2} ||Y - XBZ'||_2^2$ and $g(B) = \lambda ||B||_1$.
 
Alternatively, one can approximate the non-smooth objective function
with a twice-differentiable surrogate or recast the problem with
constraints. In the former case, unconstrained optimization approaches
like Newton's method can be used directly to minimize the suitably
chosen approximation. Two general approaches for choosing the
surrogate are replacing the non-differentiable penalty $g$ with a
fixed smooth approximation or iteratively bounding $g$ from above with
a convex function \citep{schmidt2009optimization}. ADMM is an example
of the latter approach that utilizes a dual decomposition and
augmented Lagrangians to perform constrained
optimization \citep{boyd2011distributed}.  It has also been used in
statistical applications; see for example \cite{tan2014hubs}.

The reader might be wondering why any new algorithms are needed for
this problem.  The reason is that, while mathematically equivalent to
Equation (\ref{eq:mlm}), the vectorized form in Equation (\ref{eq:vecmlm}) is
computationally cumbersome. For example, the R package
\texttt{glmnet} \citep{friedman2010regularization} fails, even for
moderate dimensions of $X$ and $Z$, because their Kronecker product
has large memory requirements.  Unlike general solvers, we utilize the
fact that the design matrix is a Kronecker product, where all of the
information is contained in two smaller matrices, and are able to
obtain a computationally efficient solution.  From a philosophical
standpoint, vectorizing the data destroys its natural structure,
making the interpretations harder (even though the numbers are same).
Finally, our approach points to how one might fit penalized
multivariate regression models for multi-dimensional (tensor-valued)
responses, i.e. sparse tensor linear models.


\section{Data}
\label{sec:data}

In this section, we give some example high-throughput datasets that
contributed to our methodological work.  We remark on the sizes of
the datasets, the nature of the biological questions, and how they
relate to our model.

\subsection{\textit{E. coli} chemical genetic screen}

\cite{nichols2011phenotypic} performed a high-throughput genetic screen
using 3,983 strains of \textit{E. coli} that carried a mutation in a
non-essential gene.  They were grown in 307 conditions representing
114 unique stresses; more than half of them were
antibiotic/antimicrobial treatments, but other conditions such as
temperature and pH were included.  Each experimental run had at least
two replicates of the same strain.  The goal was to identify
condition-gene interactions, the idea being that such interactions
would illuminate the functional role of the mutated gene. In a matrix
linear model, the row covariates are the growth conditions, the column
covariates are the mutants, and the coefficients are the
gene-condition interactions.

\subsection{Arabidopsis G$\times$E experiment}

A population of 404 \textit{Arabidopsis thaliana} recombinant lines
derived from a cross between an ecotype (strain) originating in Sweden
and an ecotype originating in Italy were grown in three consecutive
years (2009, 2010, and 2011) in both Italy and Sweden.  The main
phenotype of interest was fitness, measured by the average number of
seeds produced per plant.  The lines were genotyped at 348 markers,
and the goal was to identify genetic loci (quantitative trait loci,
QTL) contributing to fitness across sites and years (main effect QTL)
and exhibiting gene-environment interactions
\citep{aagren2017adaptive, aagren2016data}. In a matrix linear model,
the row covariates are the markers, the column covariates are the
environments, and the coefficients are the QTL (main effects and
QTL-environment interactions).

\subsection{eQTL experiment in two environments}

This study involved a population of 104 recombinant inbred lines
derived from the Tsu-1 (Tsushima, Japan) and Kas-1 (Kashmir, India)
ecotypes of \textit{Arabidopsis thaliana} \citep{lowry2013expression,
  lovell2015exploiting}. Gene expression traits were collected for
25,662 genes, and 450 markers were genotyped.  In order to identify
main effect (G) and interaction (G $\times$ E) expression quantitative
trait loci (eQTLs) with drought stress, the experiment was run on wet
and dry soil treatments with two replicates. The data structure for
this experiment is similar to the previous one, but with many more
traits.  In a matrix linear model, the row covariates are the markers,
the column covariates are the gene identity and treatment information,
and the coefficients are the eQTL (both main effects and
interactions).

\subsection{Environmental screening}

\cite{woodruff2011environmental}
analyzed biomonitoring data from the National Health and Nutritional
Examination Survey (NHANES) to characterize both individual and
multiple chemical exposures in U.S. pregnant women.  They analyzed
data for 163 chemical analytes in 12 chemical classes for subsamples
of 268 pregnant women from NHANES 2003-2004.  Most of the chemicals
were measured using mass spectrometry. In a matrix linear model, the
row covariates are the demographics of the subjects, the column
covariates are the chemical classes of the chemicals, and the
coefficients are the associations between chemical classes and demographic
variables.

\section{Computational methods}
\label{sec:comp_methods}

We outline three algorithms we used for fitting the L$_1$-penalized
model, beginning with the most stable algorithm: coordinate descent.
Next, we describe two variants of FISTA, a considerably faster, but
less stable, approach. A discussion on ADMM, which is known for its
fast convergence to an approximate solution but slow convergence to
high accuracy, follows. We conclude the section with computational and
implementation considerations.

Throughout, as in the univariate case, the intercept is omitted from
the penalty term $g$ and is thus not regularized, unless otherwise
stated.  Standardizing $X$ and $Z$ by subtracting the row means and
dividing by the row standard deviations is also recommended.

\subsection{Coordinate descent}

Cyclic coordinate descent searches for the minimum of a multivariable
function by minimizing it along one coordinate direction at a time and
cyclically iterating through each direction until convergence. When 
using the least squares loss function, it is sometimes known as the 
shooting algorithm \citep{fu1998penalized}. We
first calculate the directional derivatives along the forward and
backward directions for the coordinate direction $u_{ij}$ at each
coefficient $B_{ij}$:

\begin{equation}
\begin{split}
d_{u_{ij}} h(B) & = \lim_{\tau \to 0} \frac{h(B + \tau u_{ij}) - h(B)}{\tau} = d_{u_{ij}} f(B) + 
\begin{cases} 
\lambda, & B_{ij} \geq 0
\\ -\lambda, & B_{ij} < 0
\end{cases}
\\
d_{-u_{ij}} h(B) & = \lim_{\tau \to 0} \frac{h(B - \tau u_{ij}) - h(B)}{\tau} = d_{u_{ij}} f(B) + 
\begin{cases} 
-\lambda, & B_{ij} > 0
\\ \lambda, & B_{ij} \leq 0
\end{cases}
\end{split}
\end{equation}

This is possible because the nondifferentiable penalty $g$ has
directional derivatives along $u_{ij}$ and $-u_{ij}$. Furthermore, the
loss function $f$ is differentiable, so its forward and backward
directional derivatives are simply the positive and negative ordinary
partial derivatives stored in the gradient $\nabla f$.

\begin{equation}
\begin{split}
 d_{u_{ij}} f(B) & = \frac{\partial}{\partial B_{ij}} f(B) = \nabla f(\widehat{B}_{ij}) = - (X_{:i})'R(Z_{:j}) 
 \\
 d_{-u_{ij}} f(B) & = - \frac{\partial}{\partial B_{ij}} f(B) = - \nabla f(\widehat{B}_{ij}) = (X_{:i})'R(Z_{:j}) 
\end{split}
\end{equation}
Above, $X_{:i}$ and $Z_{:j}$ denote the $i$-th and $j$-th columns of
$X$ and $Z$ respectively; $R=Y-X \widehat{B}Z'$ is the matrix of
residuals.  Note that calculating the gradient $\nabla
f(\widehat{B}_{ij})$ involves low-dimensional matrix
multiplication. Like \cite{wu2008coordinate}, our implementation
organizes cyclic updates around the residuals, which makes calculating
$\nabla f(\widehat{B}_{ij})$ fast. For each coefficient, we compute
$\nabla f$ and then update the corresponding coefficient and residual
as follows. 

\begin{algorithm}[H]
\caption{Cyclic coordinate descent}
$\text{initialize coefficients: }\widehat{B} = 0_{p \times q}$\;
$\text{initialize previous coefficients: } \widehat{B}_{\text{prev}} = \widehat{B}$\;
$\text{calculate residuals: }R = Y - X\widehat{B}Z'$\; 
\While{not converged}{
 \For{$i = 1, ..., p \text{ and } j = 1, ..., q$}{
  $\text{calculate } \nabla f(B_{ij})$\;
  $\text{update } \widehat{B}_{ij} \gets S_{\lambda}(\widehat{B}_{ij} - \nabla f(B_{ij}))$\;
  $\text{update } R \gets R + X(\widehat{B}-\widehat{B}_{\text{prev}})Z'$\;
  $\text{update } \widehat{B}_{\text{prev},{ij}} \gets \widehat{B}_{ij}$\;
 }
}
\end{algorithm}

\noindent $S_\lambda$ is the soft-thresholding operator given as

\begin{equation}
S_{\rho}(u) = \begin{cases}
u-\rho,&u>\rho\\
0,&|u|\leq\rho\\
u+\rho,&u<-\rho, 
\end{cases}
\label{eq:soft}
\end{equation}
when $\rho= \lambda$. Our implementation uses ``warm starts'' by initializing the
coefficients at zero and computing solutions for a decreasing sequence
of $\lambda$ values. The coefficients for each subsequent $\lambda$
value are then initialized to the previous converged solutions. This
strategy is faster and leads to a more stable algorithm
\citep{friedman2010regularization}. We also take advantage of sparsity
by organizing iterations over the active set of coefficients: after
performing a full cycle through all of the coefficients, we cyclically
update only the active (nonzero) coefficients until
convergence. Another full cycle is run, and the process is repeated
until the estimates stop changing. Iterating through coefficients
randomly instead of cyclically can in practice result in faster
convergence as well, so we provide this as an option.

Note that the usage of the term ``active set'' here is related to but not the same as active set \textit{methods}, which represent a class of algorithms that iterate between updating and simultaneously optimizing a set of non-zero variables \citep{schmidt2009optimization}. This concept is also connected to the least angle regression (LARS) algorithm \citep{efron2004least}, which updates the predictor most correlated with the response by taking the largest possible step in the direction of the correlation. The process repeats until a second predictor is at least as correlated with the current residuals, and so on. LARS implementations to obtain univariate lasso coordinates exist, but have not enjoyed the same level of popularity as coordinate descent. 

\subsection{FISTA}
Coordinate descent is a very stable approach with excellent
performance for univariate L$_1$-penalized regression
\citep{wu2008coordinate}. However, it is too slow for matrix
linear models of moderately large dimensions, especially if
cross-validation is used to tune the $\lambda$ parameter. Consider
instead an iterative shrinkage-thresholding algorithm (ISTA) that
calculates the gradient at the previous coefficient estimates and
updates all of the coefficients simultaneously at each iteration
\citep{beck2009fast} as
\begin{equation}
    B^{k+1} := \text{prox}_{(\text{step} \cdot \lambda) g}(B^k - \text{step} \cdot \nabla f(B^k)).
\end{equation}
Note that the proximal operator of $g$ is simply the soft-thresholding operator $\text{prox}_{\rho g}(u) = S_{\rho} (u)$ given by Equation (\ref{eq:soft}). The updates are also multiplied by a small,
fixed step size (step) that is less than 1. While ISTA may take more iterations than
coordinate descent to converge, each iteration is faster because the
gradient can be calculated efficiently as a matrix product.

Choosing the step size requires some care, as an overly small step size
can result in slow convergence and an overly large one can lead to
divergence. A suggested approach for choosing the step size is to use
the reciprocal of the (smallest) Lipschitz constant of $\nabla f$,
given by $2 \times \{\text{maximum eigenvalue of } (Z \otimes X)'(Z \otimes X) \}$.
The maximum eigenvalue of $(Z \otimes X)'(Z \otimes X)$ is equal to the product of the
maximum eigenvalues of $Z'Z$ and $X'X$, which allows us to bypass 
computing the Kronecker product. 

Fast iterative shrinkage-thresholding algorithms (FISTAs) are an
extension of ISTA \citep{beck2009fast, nesterov1983method} that
calculate the gradient $\nabla f$ based on extrapolated coefficients
comprised of a linear combination of the coefficients at the previous
two iterations. If $\widehat{B}$ is the matrix of coefficient
estimates from the most recent iteration and
$\widehat{B}_{\text{prev}}$ is that from the second-to-last iteration,
then calculate $\nabla f$ using $A = \widehat{B}
+ \frac{k-1}{k+2}(\widehat{B} - \widehat{B}_{\text{prev}})$. This
approach takes into account the change between the coefficients in
previous iterations, leading to a ``damped oscillation'' convergence
that reduces overshooting when the local gradient is changing
quickly \citep{su2016differential}.

\begin{algorithm}
\caption{FISTA with fixed step size}
$\text{initialize coefficients: }\widehat{B} = 0_{p \times q}$\;
$\text{initialize extrapolated coefficients: } A = \widehat{B}$\;
$\text{set step size: step = } [2 \times\{\text{max eigenvalue of } (Z \otimes X)'(Z \otimes X) \} ]^{-1}$\;
$\text{set current iteration: }k = 1$\;
\While{not converged}{
 $\text{update } R \gets Y - XAZ'$\; 
 $\text{calculate } \nabla f(A)$\;
 $\text{update } \widehat{B}_{\text{prev}} \gets \widehat{B}$\;
 $\text{update } \widehat{B} \gets S_{\text{step} \cdot \lambda}(A - \text{step} \cdot \nabla f(A))$\;
 $\text{update } A \gets \widehat{B} + \frac{k-1}{k+2}(\widehat{B} - \widehat{B}_{\text{prev}})$\;
 $\text{update } k \gets k+1$\;
}
\end{algorithm}

Even faster convergence can be achieved by implementing a backtracking 
line search to find the maximum step size at each iteration, instead 
of initializing a fixed step size. The idea is that the step size 
should be small enough that the decrease in the objective function 
corresponds to the decrease expected by the gradient. First, pick an 
initial step size and choose a multiplying factor 
$0 < \gamma < 1$ with which to iteratively shrink the step size. In 
practice, we find that an initial step size of 0.01 often works well. 
At each update step, iteratively shrink the step size by multiplying 
it with $\gamma$ until it satisfies the property in equation 
(\ref{eq:stepcrit}). Then update the coefficients.

\begin{equation}\label{eq:stepcrit}
\begin{split}
& \frac{1}{2} ||Y-XBZ'||^2_2 \leq \frac{1}{2} ||Y-XAZ'||^2_2 + \langle B-A, \nabla f(A) \rangle + \frac{1}{2 \cdot \text{step}}||B-A||^2_2 
\end{split}
\end{equation}

\begin{algorithm}
\caption{FISTA with backtracking}
$\text{initialize coefficients: }\widehat{B} = 0_{p \times q}$\;
$\text{initialize extrapolated coefficients: } A = \widehat{B}$\;
$\text{initialize step size, step}$\;
$\text{choose multiplying factor } 0 < \gamma < 1$;
$\text{set current iteration: }k = 1$\;
\While{not converged}{
 \If{Eq. \ref{eq:stepcrit} not met}{
 $\text{update step} \gets \gamma \cdot \text{step}$\;
 }
 $\text{update } R \gets Y - XAZ'$\; 
 $\text{calculate } \nabla f(A)$\;
 $\text{update } \widehat{B}_{\text{prev}} \gets \widehat{B}$\;
 $\text{update } \widehat{B} \gets S_{\text{step} \cdot \lambda}(A - \text{step} \cdot \nabla f(A))$\;
 $\text{update } A \gets \widehat{B} + \frac{k-1}{k+2}(\widehat{B} - \widehat{B}_{\text{prev}})$\;
 $\text{update } k \gets k+1$\;
}
\end{algorithm}

Like coordinate descent, we implement FISTA using a path of ``warm
starts''. We note that various further refinements and extensions 
have been made for FISTA and FISTA-like algorithms in recent 
years \citep{florea2017robust, kim2018another,liang2018improving, ochs2017adaptive}. Coordinate 
descent, ISTA, and FISTA with fixed step size or
backtracking each trade off between speed and stability.

\subsection{ADMM}

Utilizing the same split of the objective function as ISTA and FISTA, the alternating direction method of multipliers (ADMM) uses the proximal operators of both $f$ and $g$. To minimize the objective function, one iterates between three updates:
\begin{align}
    B_0^{k+1} & := \text{prox}_{\rho f} (B_1^k - B_2^k) \\ 
    B_1^{k+1} & := \text{prox}_{(\lambda/\rho) g} (B_0^{k+1} +  B_2^k) \\
    B_2^{k+1} & := B_2^k + ( B_0^{k+1} - B_1^{k+1})
\end{align}
$B_0$ and $B_1$ converge to each other and to the optimal coefficient estimates $\widehat{B}$, but have slightly different properties. 

When working with the vectorized/univariate model given by Equation (\ref{eq:vecmlm}), the proximal operators of $f(\beta) = \frac{1}{2} || y - \mathbf{X}\beta ||^2_2 = \frac{1}{2} \left( y'y - 2 y' \mathbf{X}\beta + \beta' \mathbf{X}' \mathbf{X} \beta \right)$ and $g(\beta) = \lambda ||\beta||_1$ are known to be
\begin{align} 
    \text{prox}_{\rho f}(u) & = \left( \rho I + \mathbf{X}'\mathbf{X} \right)^{-1} \left( \rho u + \mathbf{X}' y \right) \label{eq:prox_f}
\end{align}
and
\begin{align} 
    \text{prox}_{\rho g}(u) & = S_{\lambda/\rho}(u). \label{eq:prox_g}
\end{align}

The soft-thresholding operator in Equation (\ref{eq:prox_g}) can conveniently be applied element-wise. However, a potential bottleneck in this scheme is the inversion of $\rho I + \mathbf{X}'\mathbf{X}$ in Equation (\ref{eq:prox_f}), so consider re-formulating $f(\beta)$ in terms of the spectral decomposition $\mathbf{X}'\mathbf{X} = Q \Lambda Q'$: 
\begin{align*}
    f(\beta) & = \frac{1}{2} \left( y'y - 2 y' \mathbf{X}\beta + \beta' \mathbf{X}' \mathbf{X} \beta \right) \\
    & = \frac{1}{2} \left( y'y - 2 y' \mathbf{X} Q Q' \beta + \beta' Q \Lambda Q' \beta \right) \\
    & = \frac{1}{2} \left[ y'y - 2 y' \mathbf{X}^* \beta^* + (\beta^*)' \Lambda \beta^* \right] = f(\beta^*), 
\end{align*}
where $\beta^* = Q' \beta$, $\beta = Q \beta^*$, and $\mathbf{X}^* = \mathbf{X} Q$. 

By applying the property that $\text{prox}_{\rho f}(u) = Q \cdot \text{prox}_{\rho f}(Q' u)$ when $Q$ is an orthogonal matrix, an equivalent update can be derived that involves element-wise division instead of matrix inversion. 
\begin{equation}
\begin{split}
    \text{prox}_{\rho f}(u) & = Q \cdot \text{prox}_{\rho f}(u^*) \\
    & = Q \left( \rho I + \Lambda \right)^{-1} \left[ \rho u^* + (\mathbf{X}^*)' y \right] \\
    & = Q \left( \rho I + \Lambda \right)^{-1} \left[ \rho Q' u + (\mathbf{X}^*)' y \right] \\
    & = Q \left[ \rho Q' u + (\mathbf{X}^*)' y \right] ./ \left[ \rho + \text{diag}(\Lambda) \right]
\end{split} \label{eq:prox_f_svd}
\end{equation}
In the above expression, $./$ denotes element-wise division and $\text{diag}(\Lambda)$ extracts the diagonal elements of $\Lambda$, namely the eigenvalues of $\mathbf{X}'\mathbf{X}$. 

To obtain the analogous proximal operators for matrix linear model updates, we return to the vectorized formulation in Equation (\ref{eq:vecmlm}) and recognize that 
\begin{align*}
    \mathbf{X}'\mathbf{X} & = (Z \otimes X)' (Z \otimes X) 
    \\
    & = (Z'Z) \otimes (X'X) \\
    & = (Q_Z \Lambda_Z Q_Z') \otimes (Q_X \Lambda_X Q_X') \\
    & = (Q_Z \otimes Q_X) (\Lambda_Z \otimes \Lambda_X) (Q_Z \otimes Q_X)', 
\end{align*}
where the third equality follows from the spectral decompositions $Z'Z = Q_Z \Lambda_Z Q_Z$ and $X'X = Q_X \Lambda_X Q_X'$. Then $Q = Q_Z \otimes Q_X$, $\Lambda = \Lambda_Z \otimes \Lambda_X$, and $\mathbf{X}^* = \mathbf{X} Q = (Z \otimes X) (Q_Z \otimes Q_X)$. Also recall that $y = \text{vec}(Y)$ and $\beta = \text{vec}(B)$, and apply Kronecker product properties to Equation (\ref{eq:prox_f_svd}) to get the final devectorized expression in Equation (\ref{eq:prox_f_mlm}): 
\begin{align}
    \text{prox}_{\rho f}(\text{vec}(U)) & = (Q_Z \otimes Q_X) \left\{ \rho (Q_Z \otimes Q_X)' \text{vec}(U) \right. \nonumber \\
    & \qquad\qquad\qquad\quad \left. + [(Z \otimes X) (Q_Z \otimes Q_X)]' \text{vec}(Y) \right\} \nonumber \\
    & \qquad ./ \left[ \rho + \text{diag}(\Lambda_Z \otimes \Lambda_X) \right] \nonumber \\
    & = (Q_Z \otimes Q_X) \text{vec} \left( \rho Q_X' U Q_Z + Q_X' X' Y Z Q_Z \right) \nonumber \\
    & \qquad ./ \left[ \rho + \text{diag}(\Lambda_Z \otimes \Lambda_X) \right] \nonumber \\
    & = (Q_Z \otimes Q_X) \text{vec} \left[ \left( \rho Q_X' U Q_Z + Y^* \right) ./ \left( \rho + L \right) \right] \nonumber \\
    & = \text{vec} \left\{ Q_X \left[ \left( \rho Q_X' U Q_Z + Y^* \right) ./ \left( \rho + L \right) \right] Q_Z' \right\} \nonumber \\
    \text{prox}_{\rho f}(U) & = Q_X \left[ \left( \rho Q_X' U Q_Z + Y^* \right) ./ \left( \rho + L \right) \right] Q_Z' \label{eq:prox_f_mlm}
\end{align}
The only necessary Kronecker product is therefore that between diagonal matrices $\Lambda_Z$ and $\Lambda_X$, a cheap calculation compared to a Kronecker product of dense matrices. One can also pre-compute $Y^* = Q_X' X' Y Z Q_Z$ and $L = \text{vec}^{-1}_{n, m} \left[ \text{diag}(\Lambda_Z \otimes \Lambda_X) \right]$. $\text{vec}^{-1}$ denotes the inverse of the vectorization operator, such that $\text{vec}^{-1}_{n, m} \left[ \text{vec} (A) \right] = A$ for all $A \in \mathbb{R}^{n \times m}$ and $\text{vec} \left[ \text{vec}^{-1}_{n, m} (a) \right] = a$ for all $a \in \mathbb{R}^{nm}$. 

When a rough solution is sufficient, ADMM can be a good approach because it is often easy to implement and converges to approximate estimates quickly. However, ADMM has been observed to be slow when a high degree of accuracy is desired. Like the choice of step size in FISTA, the choice of $\rho > 0$ to tune ADMM has consequences for the speed of convergence. To set the initial value of $\rho$, we follow the suggestion laid out by \cite{ghadimi2012optimal} for the L$_1$-regularized ADMM algorithm. When $\lambda < \text{min} \left[ \text{diag}(\Lambda_Z \otimes \Lambda_X) \right]$--- that is, when the penalty parameter $\lambda$ is less than the minimum eigenvalue of $(Z \otimes X)'(Z \otimes X)$--- we set $\rho = \text{min} \left[ \text{diag}(\Lambda_Z \otimes \Lambda_X) \right]$. When $\lambda > \text{max} \left[ \text{diag}(\Lambda_Z \otimes \Lambda_X) \right]$, we set $\rho = \lambda$; otherwise, we set $\rho = \text{max} \left[ \text{diag}(\Lambda_Z \otimes \Lambda_X) \right]$. 

Furthermore, \cite{boyd2011distributed} describe a simple approach for varying the ADMM tuning parameter such that the rate of convergence is less dependent on the initial choice of $\rho$. Define the primal residuals as $r = B_1 - B_0$ and the dual residuals $s$ as the difference between the values of $B_1$ at the previous and current iterations. At the $(k+1)$th iteration, update $\rho$ as 
\begin{align*}
    \rho^{k+1} = \begin{cases}
    \tau_{\text{incr}} \rho^k & \text{if } ||r^k||_2 > \mu ||s^k||_2 \\
    \rho^k / \tau_{\text{decr}} & \text{if } ||s^k||_2 > \mu ||r^k||_2 \\
   \rho^k & \text{otherwise}, 
    \end{cases}
\end{align*}
for some choice of parameters $\mu > 1$, $\tau_{\text{incr}} > 1$, and $\tau_{\text{decr}}$. We use the typical values, as indicated in the paper, of $\mu = 10$
and $\tau_{\text{incr}} = \tau_{\text{decr}} = 2$. If $\rho$ changes between iterations, $B_2$ must be rescaled accordingly. 

\begin{algorithm}
\caption{ADMM}
$\text{initialize all coefficients: }B_0 = B_1 = B_2 = 0_{p \times q}$\;
$\text{choose parameters for tuning } \rho: \mu > 1, \tau_{\text{incr}} > 1, \text{ and } \tau_{\text{decr}} >1$\;
$\text{obtain spectral decompositions } X'X = Q_X \Lambda_X Q_X' \text{ and } Z'Z = Q_Z \Lambda_Z Q_Z'$\;
$\text{pre-compute } Y^* = Q_X' X' Y Z Q_Z \text{ and } L = \text{vec}^{-1}_{n, m} \left[ \text{diag}(\Lambda_Z \otimes \Lambda_X) \right]$\;
$\text{initialize ADMM tuning parameter } \rho > 0 \text{ as}$\\
\uIf{$\lambda < \{\text{min eigenvalue of } (Z \otimes X)'(Z \otimes X) \}$}{
  $\text{set } \rho = \{\text{min eigenvalue of } (Z \otimes X)'(Z \otimes X) \}$\;
} \uElseIf{$\lambda > \{\text{max eigenvalue of } (Z \otimes X)'(Z \otimes X) \}$}{
  $\text{set } \rho = \lambda$\;
} \Else{
  $\text{set } \rho = \{\text{max eigenvalue of } (Z \otimes X)'(Z \otimes X) \}$\;
}  
\While{not converged}{
  $\text{set } B_{1, \text{prev}} = B_1$\;
  $\text{update } B_0 \gets Q_X \left[ \left( \rho Q_X' (B_1-B_2) Q_Z + Y^* \right) ./ \left( \rho + L \right) \right] Q_Z'$\;
  $\text{update } B_1 \gets S_{\lambda/\rho}\left(B_0+B_2 \right) $\;
  $\text{update } B_2 \gets B_2 + ( B_0 - B_1 ) $\;
  $\text{update } r \gets B_0 - B_1 $\;
  $\text{update } s \gets \rho ( B_{1, \text{prev}} - B_1) $\;
  \uIf{$||r||_2 > \mu ||s||_2$}{
    $\text{update } \rho \gets \tau_{\text{incr}} \rho$\;
    $\text{rescale } B_2 \gets B_2/\tau_{\text{incr}} $\;
  } \uElseIf{$||s||_2 > \mu ||r||_2$}{
    $\text{update } \rho \gets \rho/\tau_{\text{decr}}$\;
    $\text{rescale } B_2 \gets \tau_{\text{decr}} B_2 $\;
  } 
}
$\text{return } \widehat{B} = B_1$\;
\end{algorithm}

\subsection{Computational Considerations}

We emphasize again that while many solvers are available for the
vectorized matrix linear model given by Equation (\ref{eq:vecmlm}), this
formulation is impractical or impossible even for
moderately sized $X$ and $Z$ because their Kronecker product is too
large.  Many of the operations in our algorithms, such as matrix
multiplication and element-wise operations, are parallelizable.  With
suitable hardware and software, significant speedups are possible.  Our
implementation did not use any parallelization.

\subsubsection{Shrinkage Parameter Tuning}

To determine the optimal shrinkage/regularization parameter 
$\lambda$, $k$-fold cross-validation is recommended; a parallel 
implementation is straightforward. Various criteria can be used to 
identify optimal performance averaged across the $k$ folds, including 
mean squared error (MSE), test error, AIC, and BIC.  We used 
MSE for the analyses presented in the Results (Section \ref{sec:results}). It is also 
possible to choose a $\lambda$ based on the proportion of significant 
(nonzero) interactions desired.

\subsubsection{Software Implementation}

We implemented these algorithms using the high-level programming
language Julia \citep{bezanson2017julia}.  Julia is a relatively young
language with an active community that combines ease of prototyping
with computational speed. It features a just-in-time compiler and
strong data typing, which enable fast computation.  It is an
attractive candidate for numerical computing problems such as ours,
since one does not need to switch between multiple programming
languages for implementation, analysis, and visualization.  Julia has
built-in support for parallelization which is helpful for large-scale
analysis.

Our package's primary function, \texttt{mlmnet}, allows users to
specify the data, penalty values, and estimation algorithm.  Users can
also choose which rows and columns of $B$ should be regularized, and
if one or both of the $X$ and $Z$ intercepts should be included and/or
regularized. The package implements parallelized cross-validation for tuning
$\lambda$ and includes several functions for summarizing results.

\section{Results}
\label{sec:results}


\subsection{Simulated data with varying dimensions}
\label{sec:scaling_times}
To illustrate the speed of FISTA with backtracking and ADMM, we ran
the algorithms on simulated data while fixing the dimensions of the
multivariate response matrix and varying the dimensions of the
interaction matrix (Table \ref{tab:pq_times}), or vice versa (Table
\ref{tab:nm_times}).  The setup represents a two-way layout where
the row covariates correspond to a factor with $p$ levels and the column
covariates correspond to a factor with $q$ levels.
The data was simulated with 1/2 nonzero row and column main effects
and 1/8 nonzero interactions drawn from Normal(0, 2)
distributions. That is, to get the row main effects, we simulated a
vector of length $p$ with a random half of the entries set to zero and
the other half of the entries drawn from Normal(0, 2). The column main
effects (a vector of length $q$) and interactions (a $p \times q$
matrix) were obtained similarly. The $X$ matrix was generated by
repeatedly stacking $p \times p$ identity matrices until $n$ rows were 
reached, and analogously for $Z$.  We also included an intercept in both $X$ and $Z$ by
concatenating a column of 1s, to encode for the main effects.  Errors
were drawn from Normal(0, 3). Times are presented as averages of 100
replicates, each run over 20 $\lambda$ values. We used a dual CPU Xeon
E5-2623 v3 @ 3.00GHz processor with 125 G RAM.

Both algorithms remain fast even when scaling to greater
dimensions. Interestingly, ADMM is much faster than FISTA in cases
where $n$ and $m$ are large relative to $p$ and $q$. Its runtimes also
scale better when increasing $n$ and $m$. However, when $p$ and $q$
approach $n$ and $m$ (i.e. when $X$ and/or $Z$ are close to being
square matrices), the computational performance of ADMM suffers
greatly. $Q_X$ is $p \times p$ and $Q_Z$ is $q \times q$, so the
matrix multiplication used to transform and back-transform $B_0$ in
the ADMM updates relies heavily on the size of $p$ and $q$ rather than
$n$ and $m$.

The rate at which the runtimes increase is also not entirely
symmetrical for the two methods, both individually and relative to
each other. For example, it appears to be more computationally
expensive to increase the number of columns $q$ in $Z$ than it is to
increase the number of columns $p$ in $X$, for either method. However,
the runtimes also increase more quickly for ADMM than for FISTA when
scaling up $q$ compared to scaling up $p$.

\begin{table}[ht]
\centering
\caption{\textbf{Ratios of computation times for running FISTA with
  backtracking and ADMM on simulated data while varying $p$ and $q$ 
  (the dimensions of the interaction matrix).} Times (in minutes) 
  were obtained as averages of 100 replicates, each run over 20 
  $\lambda$ values and holding $n = m = 1200$. The raw runtimes for 
  FISTA and ADMM are reported to the left and right of the forward 
  slash, respectively (FISTA/ADMM). The cell colors indicate the magnitude of 
  the discrepancy between the two methods, based on 
  ratios of the runtimes.}
\label{tab:pq_times}
\begin{tabular}{l|rrrrr}
\hline
FISTA\big/ADMM & $q=200$ & $q=400$ & $q=600$ & $q=800$ & $q=1000$ \\
\hline
$p=200$  & \cellcolor[HTML]{333333} \textcolor[HTML]{FFFFFF}{1.20\big/0.80} & \cellcolor[HTML]{000000} \textcolor[HTML]{FFFFFF}{2.05\big/1.23} & \cellcolor[HTML]{333333} \textcolor[HTML]{FFFFFF}{2.69\big/1.71} & \cellcolor[HTML]{333333} \textcolor[HTML]{FFFFFF}{3.35\big/2.27} & \cellcolor[HTML]{666666} 3.91\big/2.89 \\
$p=400$  & \cellcolor[HTML]{333333} \textcolor[HTML]{FFFFFF}{1.29\big/0.85} & \cellcolor[HTML]{999999} 1.85\big/1.44 & \cellcolor[HTML]{999999} 2.51\big/2.10 & \cellcolor[HTML]{CCCCCC} 3.18\big/2.83 & 3.77\big/3.60 \\
$p=600$  & \cellcolor[HTML]{666666} 1.38\big/1.00 & \cellcolor[HTML]{CCCCCC} 2.05\big/1.80 & 2.72\big/2.73 & 3.47\big/3.57 & \cellcolor[HTML]{CCCCCC} 4.09\big/4.87 \\
$p=800$  & \cellcolor[HTML]{CCCCCC} 1.50\big/1.28 & 2.16\big/2.25 & \cellcolor[HTML]{CCCCCC} 2.94\big/3.34 & \cellcolor[HTML]{999999} 3.54\big/4.47 & \cellcolor[HTML]{666666} 4.14\big/6.19 \\
$p=1000$ & 1.58\big/1.59 & \cellcolor[HTML]{CCCCCC} 2.23\big/2.70 & \cellcolor[HTML]{999999} 3.09\big/3.98 & \cellcolor[HTML]{666666} 3.90\big/5.89 & \cellcolor[HTML]{666666} 4.65\big/7.90 \\
\hline
\end{tabular} 
\end{table}

\begin{table}[ht]
\centering
\caption{\textbf{Ratios of computation times for running FISTA with
  backtracking and ADMM on simulated data while varying $n$ and $m$ 
  (the dimensions of the multivariate response matrix).} Times (in 
  minutes) were obtained as averages of 100 replicates, each run 
  over 20 $\lambda$ values and holding $p = q = 400$. The raw 
  runtimes for FISTA and ADMM are reported to the left and right of 
  the forward slash, respectively (FISTA/ADMM). The cell colors indicate the 
  magnitude of the discrepancy between the two 
  methods, based on ratios of the runtimes.}
\label{tab:nm_times}
\begin{tabular}{l|rrrrr}
\hline
FISTA\big/ADMM & $m=400$ & $m=800$ & $m=1200$ & $m=1600$ & $m=2000$ \\
\hline
$n=400$  & \cellcolor[HTML]{000000} \textcolor[HTML]{FFFFFF}{0.53\big/1.35} & \cellcolor[HTML]{999999} 0.51\big/0.69 & \cellcolor[HTML]{CCCCCC} 0.64\big/0.73 & 0.83\big/0.81 & \cellcolor[HTML]{CCCCCC} 1.00\big/0.94 \\
$n=800$  & \cellcolor[HTML]{CCCCCC} 0.70\big/0.76 & \cellcolor[HTML]{CCCCCC} 0.99\big/0.91 & \cellcolor[HTML]{CCCCCC} 1.28\big/1.11 & \cellcolor[HTML]{999999} 1.55\big/1.26 & \cellcolor[HTML]{999999} 1.85\big/1.52 \\
$n=1200$ & \cellcolor[HTML]{CCCCCC} 0.99\big/0.91 & \cellcolor[HTML]{999999} 1.48\big/1.13 & \cellcolor[HTML]{999999} 1.83\big/1.46 & \cellcolor[HTML]{666666} 2.36\big/1.77 & \cellcolor[HTML]{333333} \textcolor[HTML]{FFFFFF}{3.07\big/2.07} \\
$n=1600$ & \cellcolor[HTML]{999999} 1.27\big/1.06 & \cellcolor[HTML]{999999} 1.86\big/1.44 & \cellcolor[HTML]{666666} 2.59\big/1.86 & \cellcolor[HTML]{666666} 3.24\big/2.28 & \cellcolor[HTML]{333333} \textcolor[HTML]{FFFFFF}{3.92\big/2.70} \\
$n=2000$ & \cellcolor[HTML]{999999} 1.54\big/1.23 & \cellcolor[HTML]{666666} 2.33\big/1.71 & \cellcolor[HTML]{666666} 3.13\big/2.19 & \cellcolor[HTML]{333333} \textcolor[HTML]{FFFFFF}{4.10\big/2.75} & \cellcolor[HTML]{333333} \textcolor[HTML]{FFFFFF}{4.94\big/3.32} \\
\hline
\end{tabular} 
\end{table}

\subsection{Environmental screening simulations}

We simulated data modeled after an environmental screening study
\citep{woodruff2011environmental} using mass spectrometry.  The study
measured environmental chemical concentrations in pregnant women
across various demographics in several tissues.  We simulated data
from 100 chemicals, each measured in 10 tissues for 108 women. The
tissues, chemicals, and each unique combination of tissues and
chemicals were encoded in the $Z$ matrix. That is, the $Z$ matrix is comprised of an intercept, 100 dummy variables for the chemicals, 10 dummy variables for the tissues, and $100 \times 10 = 1000$ dummy variables for the unique chemical-tissue combinations: 
\begin{equation}
    Z_{10000 \times 1111} = 
    \left[ \begin{array}{ c | c | c | c} 
        1 & & & \\
        \vdots & \text{100 chemicals} & \text{10 tissues} & \text{100 chemicals $\times$ 10 tissues} \\ 
        1 & & & \\ 
    \end{array} \right]. 
\end{equation}

\begin{figure}[!ht]
\centering
\includegraphics[width=0.8\linewidth]{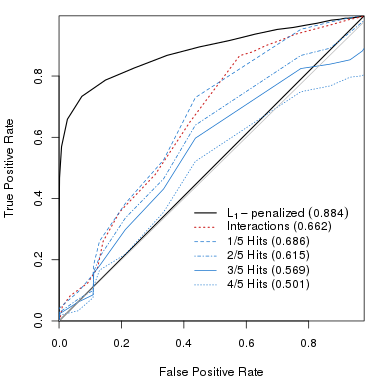}
\caption{\textbf{ROC curves for simulations comparing L$_1$-penalized
    matrix linear models to univariate linear regression for
    identifying chemical interactions in environmental screening data \citep{woodruff2011environmental}.} The AUC \citep{MESS} for each method is
  given in parentheses in the legend. L$_1$-penalized matrix 
  linear models outperforms the univariate approach.}
\label{fig:enviro_chem}
\end{figure}

We then simulated an $X$ matrix with 19 continuous demographic 
covariates drawn from the standard normal distribution, and 
included an intercept. For each tissue, 1/4 of the chemicals,
1/2 of the demographic covariates, and 1/8 of the interactions 
between chemicals/tissues and demographics, we simulated effects 
drawn from a Normal(0, 2) distribution. This was done similarly to how we generated the main effects and interactions in Section \ref{sec:scaling_times}. In this case, the demographic effects correspond to the row main effects, and the element-wise sum of the tissue and chemical effects correspond to the column effects. Errors were drawn from Normal(0, 3). We used
FISTA with backtracking to estimate the main and interaction effects. 
The receiver operating characteristic (ROC) curves in Figure 
\ref{fig:enviro_chem} compare the performance of L$_1$-penalized 
matrix linear models (MLM) to the conventional approach of running a 
univariate linear model for each chemical and tissue combination. 

\begin{itemize}
\item The solid black line plots the results for the L$_1$-penalized
  MLM. We obtained true positive rates (TPR) and false positive rates
  (FPR) by varying $\lambda$ and comparing the nonzero and zero
  interaction estimates to the true interactions.
\item The dotted red line is from running the 1000 univariate linear 
  regression models for each combination of the 100 simulated 
  chemicals and 10 simulated tissues. We obtained the adaptive
  Benjamini-Hochberg adjusted p-values \citep{benjamini2000adaptive, mutoss} for each model's coefficient 
  estimates and varied the cutoff for determining significant
  interactions. These were compared to the true interaction effects to
  calculate the TPR and FPR.
\item The blue lines of various line types offer an alternate interpretation 
  of the univariate linear models. For each chemical, there were 10 
  chemical $\times$ demographic interactions, one for each of the 10
  tissues. We flagged an interaction if least 1/5, 2/5, 3/5, or 4/5 out 
  of the 10 different p-values was below the cutoff.  A plot with 
  curves for the 10 tissues, each of which corresponds to p-values 
  from 100 different linear models, yields similar results.
\end{itemize}

Our method consistently outperforms variations of the conventional 
univariate approach. The L$_1$ penalized MLM resulted in an area under 
the curve (AUC) \citep{MESS} of 0.884; the AUC for the univariate linear regression 
interpretations was at most 0.686, which is when only one out of five 
significant univariate p-values (``hits'') is needed to detect an 
significant interaction.

\subsection{\textit{E. coli} chemical genetic screen}
A study by \cite{nichols2011phenotypic} aimed to examine the
interaction effects between 3,983 \textit{E. coli} mutant strains and
307 growth conditions. The mutant strains were taken from the Keio
single-gene deletion library \citep{baba2006construction}; essential
gene hypomorphs (C-terminally tandem-affinity
tagged \citep{butland2008esga} or specific alleles); and a small
RNA/small protein knockout library \citep{hobbs2010small}. Colony
opacity was recorded for mutant strains grown in high density on agar
plates. Six plate arrangements of mutants were used, with 1536
colonies grown per plate. In this context, a ``plate arrangement''
refers to the choice of mutants and exposures as well as their
positioning in the 1536 wells. More than half of the growth conditions
were antibiotic/antimicrobial treatments, but other types of
conditions, such as temperature and pH, were included. For this data
application, we specified a separate model for each of the six plate
arrangements. In each case, $X$ was a design matrix encoding the
growth conditions factor in the plate arrangement and $Z$ was a design
matrix encoding the mutants as a factor. We included an intercept for
the main effects in both the $X$ and $Z$ matrices.

Auxotrophs are mutant strains that have lost the ability to synthesize
a particular nutrient required for growth. Since they should
experience little to no colony growth under specific conditions where
the required nutrient is not present, we anticipate negative interactions
between auxotrophic mutants and minimal media growth conditions. While
using sparse estimates for this analysis may not be a good modeling
choice because we expect many of the interactions to be negative rather
than zero, examining auxotrophs as controls is nevertheless useful,
since the phenotype under particular conditions for a mutant strain is
typically not known.

\begin{figure}[!ht]
\centering
\includegraphics[width=0.9\linewidth]{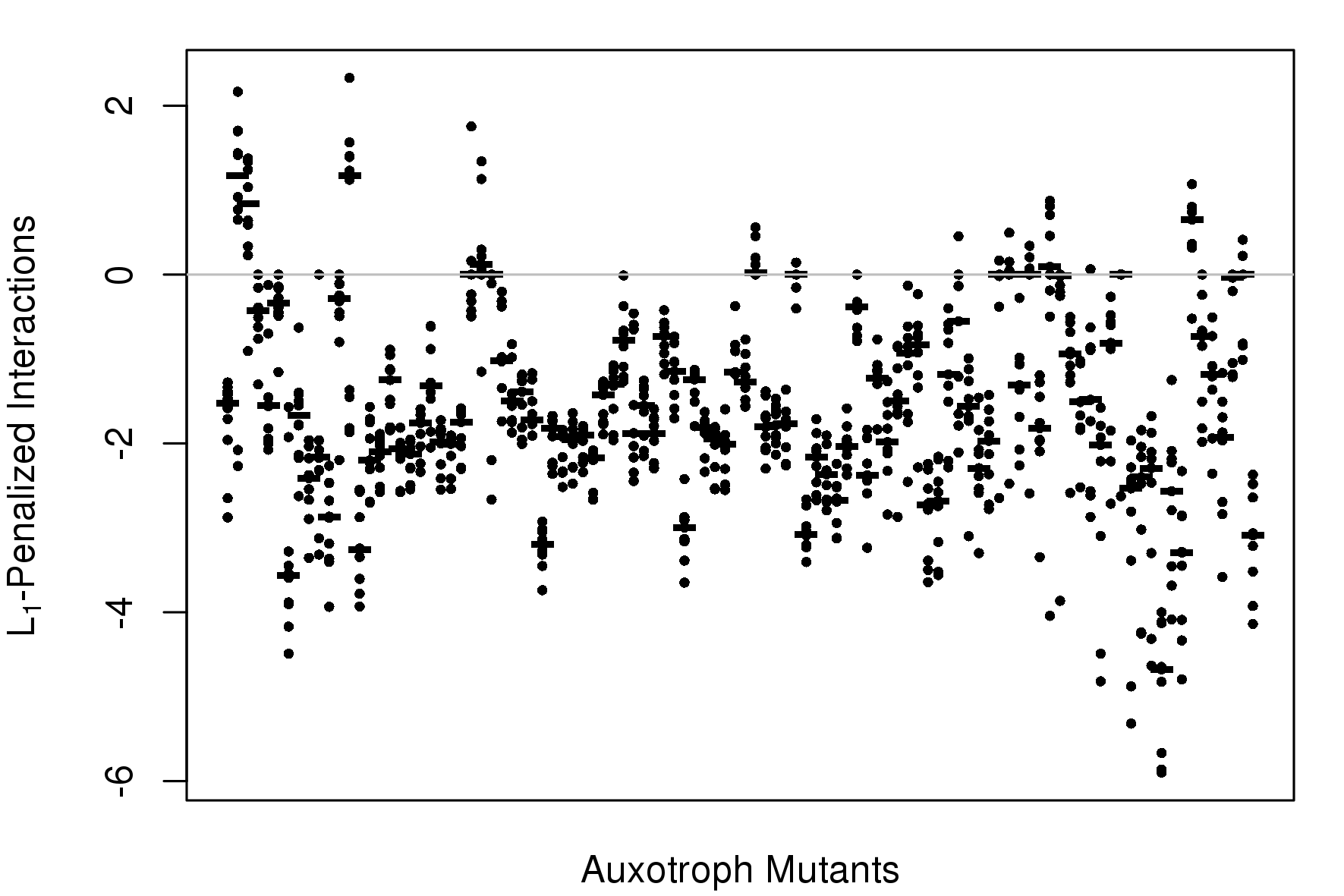}
\caption{\textbf{Distributions of sparse matrix linear model interaction estimates ($\lambda=0.46$) for auxotrophs identified by \cite{nichols2011phenotypic} over minimal media conditions in the \textit{E. coli} chemical genetic screening data.} The \cite{nichols2011phenotypic} auxotrophs are plotted along the horizontal axis. The L$_1$ penalized MLM interactions between the auxotrophs and minimal media conditions are plotted along the vertical axis, with the horizontal bars indicating the median value. Most interactions fall below zero, indicating little growth.}
\label{fig:nichols_auxo_dot}
\end{figure}

In their original analysis of the colony size data, \cite{nichols2011phenotypic} empirically identified 102 auxotrophs. Similar to what we did for the 
least-squares t-statistics (obtained by dividing the least squares
coefficient estimates by their standard errors) in
\cite{liang2019matrix}, we empirically identified auxotrophs based
on the sparse estimates. To do this, we obtained the quantiles of 
the interaction estimates for a given $\lambda$ penalty for each 
mutant strain under minimal media conditions. Mutants whose 
95\% quantile for interactions with minimal media conditions 
fell below zero were classified as auxotrophs. The lambda that 
minimizes the mean-squared error within one  cross-validation 
standard error was $\lambda=0.46$ for three of the plates and $\lambda=0.35$ for the other three plates. When setting 
$\lambda=0.46$ across all six plates, our auxotrophs had an 85\% 
overlap with the \cite{nichols2011phenotypic} auxotrophs. This is consistent with 
the 83\% overlap found in our earlier work on least-squares
t-statistics \citep{liang2019matrix}. Some of the discrepancy may 
be due to differences between analyzing colony opacity, as we did, 
and analyzing colony size, as \cite{nichols2011phenotypic} did. Figure 
\ref{fig:nichols_auxo_dot} visualizes the distributions of each 
auxotroph's sparse interactions ($\lambda=0.46$) across minimal 
media conditions. The interaction estimates are plotted as points, 
and the median for each auxotroph is plotted as a horizontal bar; 
most fall below zero. 

To compute the AUC, we took the auxotrophs identified by \cite{nichols2011phenotypic} to be the ``true'' auxotrophs. We then obtained TPRs and 
FPRs by varying cutoffs for the median minimal media interactions 
for the auxotrophs that we identified using L$_1$ penalized 
estimates and least squares t-statistics. The AUC was 0.892 for 
the L$_1$ penalized estimates and 0.884 for the least squares 
t-statistics \citep{MESS}. Supplemental Figure S1 \citep{liang2021mlmfigs} plots the ROC 
curves for the two methods, which appear very similar. So, there 
is high concordance between our empirically identified auxotrophs 
and those identified by \cite{nichols2011phenotypic}, as well as between 
our two approaches.

The auxotrophs are not expected to be sparse, but we can also consider
all of the condition $\times$ mutant interactions by using simulated
data. As we did in \cite{liang2019matrix}, we used the framework of
the $X$ and $Z$ matrices from the \textit{E. coli} data's six plate
arrangements to simulate data with 1/2 nonzero main effects and 1/4
nonzero interactions drawn from a Normal(0, 4) distribution. The
errors were independent and identically distributed from the standard
normal distribution. We then obtained the adaptive Benjamini-Hochberg
adjusted \citep{benjamini2000adaptive, mutoss} permutation p-values
from the least squares estimates and the L$_1$-penalized estimates for
50 $\lambda$ values. To compare the results for each plate
arrangement, we considered the AUCs and plotted the ROC curves. For
the least squares approach, we obtained TPRs and FPRs by varying
p-value cutoffs to determine which adjusted p-values correspond to
significant (nonzero) interactions. For the L$_1$-penalized solutions,
we obtained TPRs and FPRs by varying $\lambda$ and comparing the
nonzero and zero interaction estimates to the true interactions.

\begin{figure}[!ht]
\centering
\includegraphics[width=0.8\linewidth]{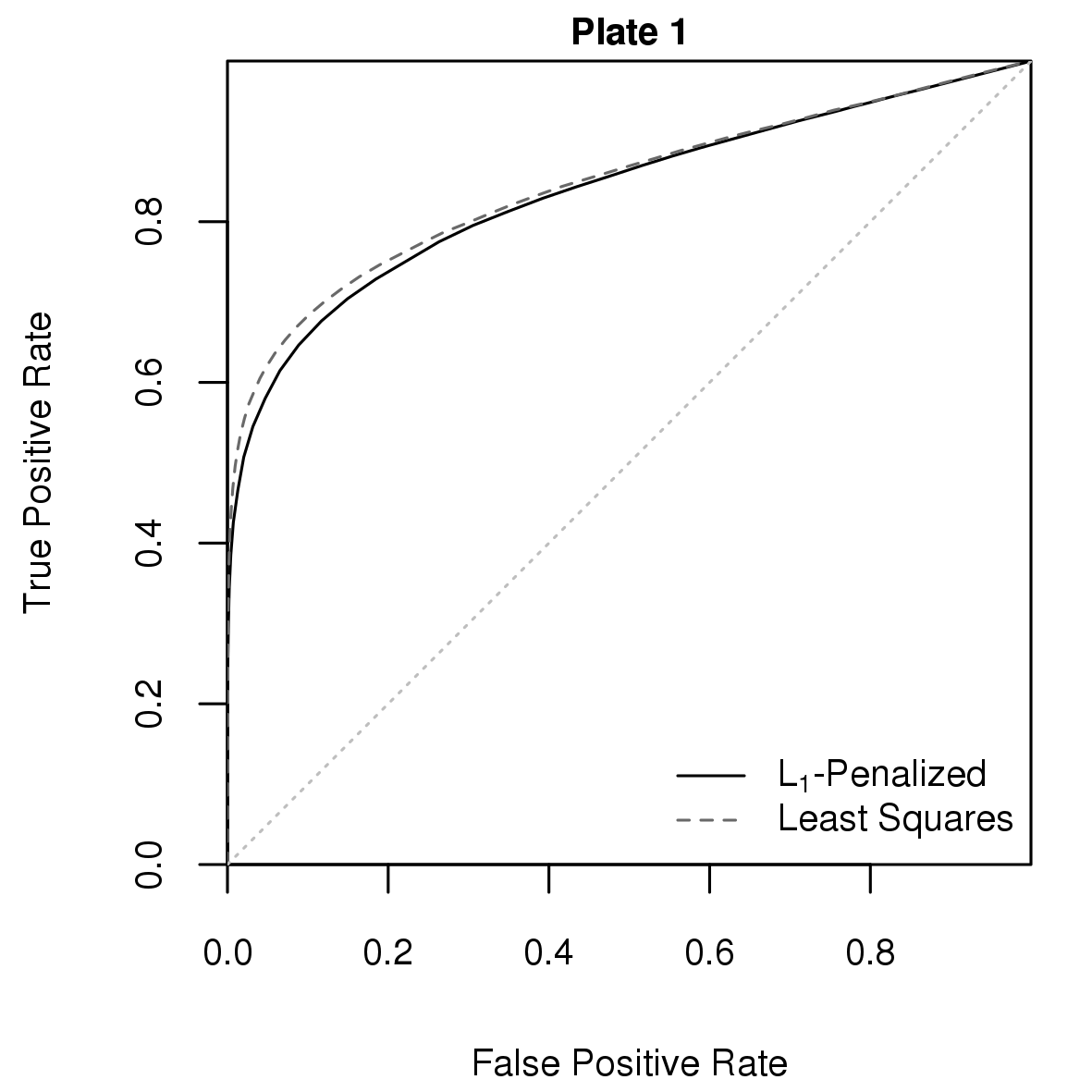}
\caption{\textbf{ROC curves comparing least squares to L$_1$-penalized estimates applied to data simulated using framework of first plate arrangement in the \textit{E. coli} chemical genetic screening data \citep{nichols2011phenotypic}.} The two methods perform very similarly, with the least squares t-statistics (AUC \citep{MESS} of 0.845) performing slightly better than the L$_1$-penalized solutions (AUC of 0.835).}
\label{fig:nichols_sim_p1_ROC}
\end{figure}

The AUCs were very similar between the two methods for all plates
(Table \ref{tab:nichols_sim_AUC}), but were consistently higher for
the least squares approach. Figure \ref{fig:nichols_sim_p1_ROC} plots
the ROC curves for the first plate arrangement, in which it is
apparent that the two curves are nearly identical in trajectory. The
ROC curves for the remaining five plates, which look quite similar to
those in Figure \ref{fig:nichols_sim_p1_ROC}, are shown in
Supplemental Figure S2 \citep{liang2021mlmfigs}. In this situation, it appears that it is both
simpler and more effective to use the closed-form least squares
estimates, rather than attempting regularization.

\begin{table}[ht]
    \centering
    \begin{tabular}{l|rrrrrr}
        \hline
        Plate           &        1 &       2 &       3 &       4 &      5 &     6 \\
        \hline
        L$_1$-Penalized & 0.835   & 0.830   & 0.778   & 0.843   & 0.838   & 0.843 \\
        Least Squares   & 0.845   & 0.843   & 0.853   & 0.852   & 0.847   & 0.854 \\
        \hline
    \end{tabular}
    \caption{\textbf{Area under the curve \citep{MESS} for simulations based on each of the six plate arrangements in the \textit{E. coli} chemical genetic screening data \citep{nichols2011phenotypic}.} The least squares matrix linear models are similar to, but consistently outperform, the L$_1$-penalized solutions in each of the six cases.}
    \label{tab:nichols_sim_AUC}
\end{table}

\subsection{\textit{Arabidopsis} G$\times$E experiment}


\cite{aagren2013genetic} studied 404 \textit{Arabidopsis thaliana}
recombinant inbred lines derived by crossing two ecotypes originating
in Italy and Sweden. The plants were grown in six environments:
in two sites (Italy and Sweden) measured over three years
(2009-2011). The investigators genotyped 348 markers with the goal of
mapping quantitative trait loci (QTL) to explain genetic mechanisms of
fitness adaptation to local environments \citep{aagren2017adaptive,
  aagren2016data}.  We ran L$_1$-penalized MLMs to determine
significant interactions between the markers and each of the two sites
and six environments. The 348 markers were encoded as dummy variables
in the $X$ matrix, which also contained an intercept. The $Z$ matrix
was comprised of a contrast between the two sites (Italy and Sweden),
representing the QTL-site interactions, and a regularized intercept,
representing main effect QTL:
\begin{equation}
    Z_{6 \times 2}' = \begin{bmatrix} 
        1 & 1 & 1 & 1 & 1 & 1 \\
        1 & -1 & 1 & -1 & 1 & -1
    \end{bmatrix}. 
\end{equation} 
We used fruit production per seedling as the response data, and
only considered the 390 lines with complete response data for all 6
environments. Data pre-processing was performed in R \citep{R} with
the help of the \texttt{R/qtl} package \citep{broman2003r}.

\begin{figure}[!ht]
\centering
\includegraphics[width=0.9\linewidth]{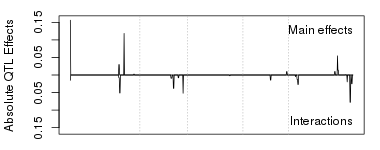}
\caption{\textbf{Absolute QTL effects plotted against marker position
    for the \cite{aagren2013genetic} \textit{Arabidopsis} data.} The
  absolute main QTL effects are plotted above the x-axis and the
  absolute QTL-site (Italy vs. Sweden) interactions are plotted below
  the x-axis.  Dotted vertical reference lines separate the five
  chromosomes.  We see main effect QTLs on chromosomes 1, 4, and 5;
  and interaction effects on chromosomes 1 2, 4, and 5.}
\label{fig:agren_site_allQTL}
\end{figure}

We performed 10-fold cross-validation with MSE as the criterion to 
determine an optimal $\lambda$ penalty size of 6.2.
Figure \ref{fig:agren_site_allQTL} plots the absolute main QTL effects
(above the x-axis) and the absolute QTL-site (Italy vs.  Sweden)
interactions (below the x-axis) against marker position on the five
chromosomes. Dotted vertical reference lines separate the chromosomes,
and the peaks correspond to loci with significant, nonzero
interactions. Several peaks are apparent on chromosomes 1, 2, 4, and
5. These results are largely aligned with the significant QTL found by
\cite{aagren2013genetic}, although it is notable that they identified
significant QTL on chromosome 3, and we did not find any. Our approach
has the advantage of being able to quickly analyze all six
environments simultaneously by encoding the $Z$ matrix with the site
information.

Table \ref{tab:compare_times} compares the times for running our
implementations of the different algorithms for obtaining L$_1$-penalized estimates on this dataset. We averaged 100 replicates
obtained from a dual CPU Xeon E5-2623 v3 @ 3.00GHz processor with 
125 G RAM. For this moderately-sized data and 50 $\lambda$ values, 
the L$_1$-penalized coefficients can be computed within a few 
minutes using FISTA and ADMM. In this case, ADMM is the fastest, 
and FISTA with backtracking is slightly slower than FISTA with a 
fixed step size. Both coordinate descent algorithms take several 
times longer, with average runtimes of over 10 minutes. Here, cyclic 
coordinate descent is faster than random coordinate descent, but 
iterating over random directions may be faster in other scenarios. 
ISTA is notable for taking more than an hour to complete, which may 
be in part due to this particular dataset and in part due to the 
choice of fixed step size. Based on our experience, while it is 
unusual to observe such an egregiously slow performance from ISTA, 
it is consistently slower than FISTA. 

\begin{table}[ht]
\centering
\caption{\textbf{Computation time (minutes) to obtain interactions
  for the \cite{aagren2013genetic} \textit{Arabidopsis} data for 
  cyclic coordinate descent, random coordinate descent, ISTA with 
  fixed step size, FISTA with fixed step size, FISTA with 
  backtracking, and ADMM.} Times were obtained as averages of 100 
  replicates, each run over 50 $\lambda$ values.} 
\label{tab:compare_times}
\begin{tabular}{lr}
\hline
Algorithm & Time (min) \\
\hline
Coordinate descent (cyclic) &  10.56 \\
Coordinate descent (random) &  10.97 \\
ISTA (fixed step size)      &   67.01 \\
FISTA (fixed step size)     &   1.39 \\
FISTA (backtracking)        &   1.50 \\
ADMM                        &   1.02 \\
\hline
\end{tabular}
\end{table}

\subsection{eQTL experiment in two environments}

\cite{lowry2013expression} examined the regulation and evolution of 
gene expression by considering drought stress. This expression 
quantitative trait locus (eQTL) mapping experiment studied
104 individuals from the Tsu-1 (Tsushima, Japan) $\times$ Kas-1
(Kashmir, India) recombinant inbred line population of
\textit{Arabidopsis thaliana}. It was conducted across wet and dry 
soil treatments with two replicates. Gene expression phenotypes were
collected for 25,662 genes, and 450 markers were genotyped
\citep{lowry2013expression, lovell2015exploiting}. The goal 
was to identify main effect (G) and interaction (G $\times$ E) eQTLs 
for the environmental conditions. Here, the $X$ matrix encoded dummy 
variables for the 450 markers, an additional treatment contrast 
encoding cytoplasm, and the intercept. The $Z$ matrix, which encoded 
for main effects and drought treatment interactions for the 25,662 
expression phenotypes, can be expressed as 

\begin{equation}
Z_{51324 \times 51324} = I_{25662} 
\otimes 
\begin{bmatrix} 
    1 & 1 \\
    1 & -1 
\end{bmatrix}. 
\end{equation}
Data pre-processing was performed in R \citep{R} with the help of
the \texttt{R/data.table} \citep{datatable} and \texttt{R/qtl}
packages \citep{broman2003r}.


It took 6.96 hours to run the FISTA algorithm for a path of 16 
$\lambda$ penalties and 13.06 hours for ADMM (ADMM is likely slow because $Z$ is a huge square matrix). We used a dual CPU Xeon 
E5-2623 v3 @ 3.00GHz processor with 125 G RAM. However, applying 
the \texttt{R/qtl} package's \texttt{stepwiseqtl} function
\citep{broman2003r} one by one for each phenotype is estimated to 
take many times as long, at 80.00 hours. This estimate was obtained by
running \texttt{stepwiseqtl} on 100 random phenotypes and
extrapolating the resulting time, averaged over 10 runs, to the full
set of 51,324 phenotypes. The large $Z$ matrix also showcases another
advantage of using FISTA with backtracking. Performing the spectral
decomposition needed to compute the fixed step size for FISTA or to
update $B_0$ in ADMM easily exceeds memory limits for a typical
computer. We were only able to run ADMM in this case because the $Z$
matrix has a special structure such that the eigenvectors form an
identity matrix. Using a backtracking line search sidesteps these
dilemmas altogether.

\begin{figure}[!ht]
\centering
\includegraphics[width=0.9\linewidth]{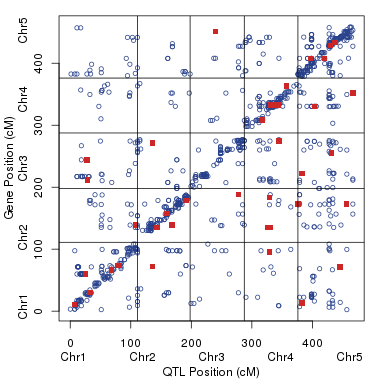}
\caption{\textbf{Distribution of eQTL across genome in the \cite{lowry2013expression} dataset.} Main effects are
  shown in open blue circles and G $\times$ E interactions in 
  solid red squares.}
\label{fig:lowry_gene}
\end{figure}

Figure \ref{fig:lowry_gene}, which reproduces Figure 2 in 
\cite{lowry2013expression} using our results, visually summarizes the 
main effect and interaction eQTLs identified by our method when 
$\lambda = 1.73$. Our method was able to detect many of the same main 
effects and G $\times$ E effects.

\begin{figure}[ht]
\centering
\includegraphics[width=0.9\linewidth]{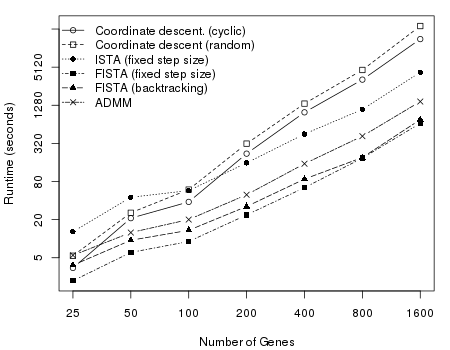}
\caption{\textbf{Computation time (seconds) to estimate interactions
  for subsets of the \cite{lowry2013expression} data, plotted 
  against the number of genes taken in the subset.} Random subsets 
  were taken of 25, 50, 100, 200, 400, 800, and 1600 genes. Runtimes 
  are shown for cyclic coordinate descent, random coordinate descent, 
  ISTA with fixed step size, FISTA with fixed step size, FISTA with 
  backtracking, and ADMM. Times were obtained as averages of 100 
  replicates, each run over 16 $\lambda$ values. Both axes are shown 
  on a log scale with base 2.}
\label{fig:lowry_times}
\end{figure}

While speed was perhaps not a huge practical concern for the \cite{aagren2013genetic} data because all of the algorithms finished within minutes, the large number of phenotypes in this dataset makes performing L$_1$-penalized MLMs a much more computationally-intensive endeavor. We compared the runtimes for the different algorithms applied to subsets of the 25,662 genes used to record gene expression phenotypes. To do this, we took random subsets of 25, 50, 100, 200, 400, 800, and 1600 genes. Note that because there were both wet and dry soil environments, the number of genes is equal to half the number of columns of $Y$ and the number of rows and columns of $Z$, i.e. if we use a subset of 20 genes, $m = q = 2 \times 20 = 40$. We used 16 $\lambda$ values for each run and averaged 100 replicates obtained from a dual CPU Xeon E5-2623 v3 @ 3.00GHz processor with 125 G RAM. The results are plotted in Figure \ref{fig:lowry_times} and reported in Table \ref{tab:lowry_times}. 

\begin{table}[ht]
\centering
\caption{\textbf{Computation time (minutes) to obtain interactions
  for subsets of the \cite{lowry2013expression} data for cyclic 
  coordinate descent, random coordinate descent, ISTA with fixed step 
  size, FISTA with fixed step size, FISTA with backtracking, and ADMM.} 
  Random subsets were taken of 25, 50, 100, 200, 400, 800, and 1600 
  genes. Times were obtained as averages of 100 replicates, each run 
  over 16 $\lambda$ values.} 
\label{tab:lowry_times}
\begin{tabular}{lrrrrrrr}
\hline
  & 25 & 50 & 100 & 200 & 400 & 800 & 1600\\
\hline
Coord. desc. (cyclic) & 0.06 & 0.35 & 0.64 & 3.67 & 16.49 & 54.20 & 236.59\\
\hline
Coord. desc. (random) & 0.09 & 0.43 & 0.99 & 5.29 & 22.65 & 77.02 & 380.74\\
\hline
ISTA (fixed step size) & 0.21 & 0.75 & 0.96 & 2.62 & 7.50 & 18.34 & 70.93\\
\hline
FISTA (fixed step size) & 0.04 & 0.10 & 0.15 & 0.40 & 1.07 & 3.13 & 11.07\\
\hline
FISTA (backtracking) & 0.06 & 0.16 & 0.23 & 0.54 & 1.46 & 3.19 & 12.66\\
\hline
ADMM & 0.09 & 0.21 & 0.33 & 0.82 & 2.54 & 6.94 & 24.43\\
\hline
\end{tabular}
\end{table}

When only a subset of 25 genes are included in the dataset, estimating the interactions can be done in a matter of seconds for all six algorithms. In such cases, there may not be much of a practical advantage in using an algorithm that requires specifying a good step size or other tuning parameters, compared to the more stable coordinate descent. However, as the number of genes included increases, the gap between the amount of time needed by coordinate descent vs. the amount of time needed by FISTA and ADMM grows. (As the number of genes and the size of $q$ grows, it also becomes apparent that ADMM is slower compared to FISTA in this case.) When 1600 genes are randomly subsetted, FISTA and ADMM still finish within 10 to 25 minutes, compared to the several hours required by coordinate descent. Extrapolating from this, it is not unreasonable to conclude that it would take on the order of days or even weeks to run coordinate descent on the full 25,662 genes in the dataset. This is not very viable, especially when faster alternatives exist. 

Cyclic coordinate descent is consistently faster than random coordinate descent for a given number of subsetted genes, which was also observed for the \cite{aagren2013genetic} data. There is some additional overhead associated with randomly selecting and accessing the coefficients. Moreover, because the QTLs per gene are more-or-less independent, iterating through the coefficients randomly may not be advantageous compared to doing it cyclically. 

If the number of subsetted genes is small, ISTA is the slowest method. However, ISTA quickly overtakes the coordinate descent algorithms as the number of genes increases. The primary difference between ISTA and coordinate descent is updating the gradient simultaneously for all of the coefficients using matrix multiplication vs. updating the coefficients one at a time. It appears that ISTA does not enjoy efficiencies from matrix multiplication when the dimensions of the coefficient matrix $B$ are relatively small, like they are when only a few genes are subsetted or in the \cite{aagren2013genetic} data. Likewise, FISTA with fixed step size is also faster than FISTA with backtracking, but the discrepancy shrinks as the number of subsetted genes increases. If the dimensions are small, the additional overhead of performing a backtracking line search costs more than the speed gained by having a dynamically updated step size.

\section{Discussion}
\label{sec:discussion}

We have developed a fast fitting procedure for sparse, L$_1$-penalized 
matrix linear models and demonstrated their use for several high-throughput
data problems.  This approach opens up analytic options for many studies using high-throughput data.  Our method takes advantage of the structure
of matrix linear models to speed up existing computational algorithms 
that cannot be feasibly run in the vectorized, univariate setting. 
Analysis of simulated and several previously-analyzed datasets 
illustrate our method's applicability. We note that as in the case of 
univariate linear regression models, whether or not to use L$_1$ 
penalized matrix linear models is a decision that should be dictated by 
the scientific goals of the study.

The choice between coordinate descent, the various flavors of (F)ISTA, 
and ADMM is largely a trade-off between speed and
stability. Coordinate descent is a reasonably fast approach for
computing L$_1$-penalized estimates for univariate linear models, but
is too slow for our multivariate scenario. Instead, we turned to the
latter two options, combined with the exploitation of the matrix
properties and sparsity of our model. The relative speed of ADMM
compared to FISTA may depend on the relative dimension sizes 
of the data. When the number of interactions (implied by the sizes of 
$p$ and $q$) is low relative to the dimensions of the response data 
(implied by the sizes of $n$ and $m$), ADMM is likely to be the fastest 
option. However, we note that the from-scratch implementation of
ADMM is quite straightforward, compared to FISTA when incorporating a
backtracking line search. 

Our work demonstrates the feasibility of fitting matrix linear models
with moderately large dimensions. It can be extended in several
promising directions that will further broaden the applicability of this
class of models.  First, we can extend fitting models with different
loss ($f$) and penalty ($g$) functions. For example, we can fit the
elastic-net method \citep{zou2005regularization} by changing the
penalty function, adding an L$_2$ penalty to the L$_1$ penalty. We 
can also make the solution less sensitive to outliers by using a robust 
loss function, such as Huber's loss function, instead of the squared 
error loss. Our current implementation allows users to specify which 
row and column effects (including intercepts) to regularize. If we 
incorporate the elastic-net method, a natural extension would be 
to also allow users to separately decide which penalty terms (if any) 
should be applied to each of the coefficients 

Another direction would be the development of confidence intervals and
hypothesis tests in this setting to complement our estimation
algorithms.  There has been some recent promising work in this
direction
\citep{javanmard2014confidence,reid2016lasso} for L$_1$ penalized 
univariate regression models.  Since the matrix linear model can be
vectorized to a univariate linear model, these results may be expected
to apply to the matrix case.

A third direction would be to extend the models to multi-dimensional
(tensor-valued) responses.  In Equation (\ref{eq:vecmlm}), the design matrix
is a Kronecker product of two matrices. Through iterative
vectorization, this model can be extended to more than two matrices to
handle a multi-dimensional $\mathscr{Y}$ response tensor. Consider a
3-way tensor $\mathscr{Y}_{n \times m \times l}$ with rows, columns,
and pages (horizontal, lateral, and frontal slices) annotated by
$X_{n \times p}$, $Z_{m \times q}$, and $W_{l \times r}$,
respectively. The goal is to estimate a 3-way tensor of coefficients,
$\mathscr{B}_{p \times q \times r}$.  $\mathscr{Y}$ can be matricized
into an $(nm) \times l$ matrix $Y^*$. To do this, each $n \times m$
frontal slice is vectorized and the resulting column vectors are laid
out into $l$ columns. One can similarly define $B^*_{(pq) \times r}$,
the matricized version of $\mathscr{B}_{p \times q \times r}$, and
$E^*_{(nm) \times l}$, the matricized version of the errors
$\mathscr{E}_{n \times m \times l}$. Let $X^*_{(nm) \times (pq)} =
(Z_{m \times q} \otimes X_{n \times p})$. Then the tensor linear model
can be written in the form of Equation (\ref{eq:mlm})
\begin{align}
    Y^* = X^* B^* W' + E^* 
\end{align}
and further reduced to the form of Equation (\ref{eq:vecmlm})
\begin{align}
    \text{vec}(Y^*) & = (W \otimes X^*) \text{vec}(B^*) + \text{vec}(E^*) \\
    \Longleftrightarrow \quad \text{vec}(\mathscr{Y}) & = (W \otimes Z \otimes X) \text{vec}(\mathscr{B}) + \text{vec}(\mathscr{E}). 
\end{align}

In summation notation, the model can be expressed as
\begin{equation}
    y_{ijk} = \sum_{s=1}^p \sum_{t=1}^q \sum_{u=1}^r x_{is} \;
    z_{jt} \; w_{ku} \; b_{stu} + e_{ijk}.
\end{equation}

The extensions for models involving higher-dimensional tensors follow 
analogously with additional iterative vectorization. These models might 
be attractive for handling, for example, time series high-throughput 
data or 3-D imaging data.  Further work is needed to explore the 
performance, scalability, and stability of the fitting algorithms for 
tensor linear models.

A fourth direction would be to consider faster implementations,
especially those with multi-threaded, distributed, or GPU computing
options, which have had recent success in machine learning.  While we
have used some of those ideas in our implementation, there is room for
considerable improvement. Reducing the computational burden would be
important for tensor linear models.

Our algorithms have been
implemented in the Julia \citep{bezanson2017julia} programming
language and are available
at \\ \href{https://github.com/senresearch/MatrixLMnet.jl}{\tt{https://github.com/senresearch/MatrixLMnet.jl}} \citep{liang2021mlmpackage}. The Julia and R \citep{R} code for reproducing the analysis and generating the figures in this paper can be found at \\ \href{https://github.com/senresearch/mlm\_l1\_supplement}{{\tt https://github.com/senresearch/mlm\_l1\_supplement}} \citep{liang2021mlmrepo}.

\begin{supplement}[id=suppA]
  \sname{Supplement A}
  \stitle{Supplemental figures}
  \slink[doi]{10.1214/00-AOASXXXXSUPP}
  \sdatatype{.pdf}
  \sdescription{Additional figures from analyzing the \textit{E. coli} chemical genetic screening data.}
\end{supplement}

\begin{supplement}[id=suppB]
  \sname{Supplement B}
  \stitle{Julia implementation for L$_1$-penalized matrix linear models}
  \slink[doi]{10.1214/00-AOASXXXXSUPP}
  \sdatatype{.zip}
  \sdescription{MatrixLMnet Julia package for estimating L$_1$-penalized matrix linear models; most up-to-date version available at \href{https://github.com/senresearch/MatrixLMnet.jl}{\tt{https://github.com/senresearch/MatrixLMnet.jl}}.}
\end{supplement}

\begin{supplement}[id=suppC]
  \sname{Supplement C}
  \stitle{Code to reproduce paper analysis}
  \slink[doi]{10.1214/00-AOASXXXXSUPP}
  \sdatatype{.zip}
  \sdescription{Repository  with code to perform analysis and 
generate figures in paper, also available at \\ \href{https://github.com/senresearch/mlm\_l1\_supplement}{{\tt https://github.com/senresearch/mlm\_l1\_supplement}}.}
\end{supplement}

\bibliographystyle{imsart-nameyear}
\bibliography{refs}

\begin{thebibliography}{46}

\bibitem[\protect\citeauthoryear{{\AA}gren et~al.}{2013}]{aagren2013genetic}
\begin{barticle}[author]
\bauthor{\bsnm{{\AA}gren},~\bfnm{Jon}\binits{J.}},
  \bauthor{\bsnm{Oakley},~\bfnm{Christopher~G.}\binits{C.~G.}},
  \bauthor{\bsnm{McKay},~\bfnm{John~K.}\binits{J.~K.}},
  \bauthor{\bsnm{Lovell},~\bfnm{John~T.}\binits{J.~T.}} \AND
  \bauthor{\bsnm{Schemske},~\bfnm{Douglas~W.}\binits{D.~W.}}
(\byear{2013}).
\btitle{Genetic mapping of adaptation reveals fitness tradeoffs in
  {A}rabidopsis thaliana}.
\bjournal{Proceedings of the National Academy of Sciences}
\bvolume{110}
\bpages{21077--21082}.
\end{barticle}
\endbibitem

\bibitem[\protect\citeauthoryear{{\AA}gren et~al.}{2016}]{aagren2016data}
\begin{bmisc}[author]
\bauthor{\bsnm{{\AA}gren},~\bfnm{Jon}\binits{J.}},
  \bauthor{\bsnm{Oakley},~\bfnm{Christopher~G}\binits{C.~G.}},
  \bauthor{\bsnm{Lundemo},~\bfnm{Sverre}\binits{S.}} \AND
  \bauthor{\bsnm{Schemske},~\bfnm{Douglas~W}\binits{D.~W.}}
(\byear{2016}).
\btitle{Adaptive divergence in flowering time among natural populations of
  Arabidopsis thaliana: estimates of selection and QTL mapping}.
\bnote{Data from: Dryad Digital Repository.
  \url{https://doi.org/10.5061/dryad.77971}}.
\end{bmisc}
\endbibitem

\bibitem[\protect\citeauthoryear{{\AA}gren et~al.}{2017}]{aagren2017adaptive}
\begin{barticle}[author]
\bauthor{\bsnm{{\AA}gren},~\bfnm{Jon}\binits{J.}},
  \bauthor{\bsnm{Oakley},~\bfnm{Christopher~G}\binits{C.~G.}},
  \bauthor{\bsnm{Lundemo},~\bfnm{Sverre}\binits{S.}} \AND
  \bauthor{\bsnm{Schemske},~\bfnm{Douglas~W}\binits{D.~W.}}
(\byear{2017}).
\btitle{Adaptive divergence in flowering time among natural populations of
  Arabidopsis thaliana: Estimates of selection and QTL mapping}.
\bjournal{Evolution}
\bvolume{71}
\bpages{550--564}.
\end{barticle}
\endbibitem

\bibitem[\protect\citeauthoryear{Baba et~al.}{2006}]{baba2006construction}
\begin{barticle}[author]
\bauthor{\bsnm{Baba},~\bfnm{Tomoya}\binits{T.}},
  \bauthor{\bsnm{Ara},~\bfnm{Takeshi}\binits{T.}},
  \bauthor{\bsnm{Hasegawa},~\bfnm{Miki}\binits{M.}},
  \bauthor{\bsnm{Takai},~\bfnm{Yuki}\binits{Y.}},
  \bauthor{\bsnm{Okumura},~\bfnm{Yoshiko}\binits{Y.}},
  \bauthor{\bsnm{Baba},~\bfnm{Miki}\binits{M.}},
  \bauthor{\bsnm{Datsenko},~\bfnm{Kirill~A}\binits{K.~A.}},
  \bauthor{\bsnm{Tomita},~\bfnm{Masaru}\binits{M.}},
  \bauthor{\bsnm{Wanner},~\bfnm{Barry~L}\binits{B.~L.}} \AND
  \bauthor{\bsnm{Mori},~\bfnm{Hirotada}\binits{H.}}
(\byear{2006}).
\btitle{Construction of Escherichia coli K-12 in-frame, single-gene knockout
  mutants: the Keio collection}.
\bjournal{Molecular systems biology}
\bvolume{2}.
\end{barticle}
\endbibitem

\bibitem[\protect\citeauthoryear{Beck and Teboulle}{2009}]{beck2009fast}
\begin{barticle}[author]
\bauthor{\bsnm{Beck},~\bfnm{Amir}\binits{A.}} \AND
  \bauthor{\bsnm{Teboulle},~\bfnm{Marc}\binits{M.}}
(\byear{2009}).
\btitle{A fast iterative shrinkage-thresholding algorithm for linear inverse
  problems}.
\bjournal{SIAM journal on imaging sciences}
\bvolume{2}
\bpages{183--202}.
\end{barticle}
\endbibitem

\bibitem[\protect\citeauthoryear{Benjamini and
  Hochberg}{2000}]{benjamini2000adaptive}
\begin{barticle}[author]
\bauthor{\bsnm{Benjamini},~\bfnm{Yoav}\binits{Y.}} \AND
  \bauthor{\bsnm{Hochberg},~\bfnm{Yosef}\binits{Y.}}
(\byear{2000}).
\btitle{On the adaptive control of the false discovery rate in multiple testing
  with independent statistics}.
\bjournal{Journal of educational and Behavioral Statistics}
\bvolume{25}
\bpages{60--83}.
\end{barticle}
\endbibitem

\bibitem[\protect\citeauthoryear{Bezanson et~al.}{2017}]{bezanson2017julia}
\begin{barticle}[author]
\bauthor{\bsnm{Bezanson},~\bfnm{Jeff}\binits{J.}},
  \bauthor{\bsnm{Edelman},~\bfnm{Alan}\binits{A.}},
  \bauthor{\bsnm{Karpinski},~\bfnm{Stefan}\binits{S.}} \AND
  \bauthor{\bsnm{Shah},~\bfnm{Viral~B}\binits{V.~B.}}
(\byear{2017}).
\btitle{Julia: A fresh approach to numerical computing}.
\bjournal{SIAM review}
\bvolume{59}
\bpages{65--98}.
\end{barticle}
\endbibitem

\bibitem[\protect\citeauthoryear{Boyd et~al.}{2011}]{boyd2011distributed}
\begin{barticle}[author]
\bauthor{\bsnm{Boyd},~\bfnm{Stephen}\binits{S.}},
  \bauthor{\bsnm{Parikh},~\bfnm{Neal}\binits{N.}},
  \bauthor{\bsnm{Chu},~\bfnm{Eric}\binits{E.}},
  \bauthor{\bsnm{Peleato},~\bfnm{Borja}\binits{B.}},
  \bauthor{\bsnm{Eckstein},~\bfnm{Jonathan}\binits{J.}} \betal{et~al.}
(\byear{2011}).
\btitle{Distributed optimization and statistical learning via the alternating
  direction method of multipliers}.
\bjournal{Foundations and Trends{\textregistered} in Machine learning}
\bvolume{3}
\bpages{1--122}.
\end{barticle}
\endbibitem

\bibitem[\protect\citeauthoryear{Broman et~al.}{2003}]{broman2003r}
\begin{barticle}[author]
\bauthor{\bsnm{Broman},~\bfnm{Karl~W}\binits{K.~W.}},
  \bauthor{\bsnm{Wu},~\bfnm{Hao}\binits{H.}},
  \bauthor{\bsnm{Sen},~\bfnm{{\'S}aunak}\binits{{\'S}.}} \AND
  \bauthor{\bsnm{Churchill},~\bfnm{Gary~A}\binits{G.~A.}}
(\byear{2003}).
\btitle{R/qtl: QTL mapping in experimental crosses}.
\bjournal{Bioinformatics}
\bvolume{19}
\bpages{889--890}.
\end{barticle}
\endbibitem

\bibitem[\protect\citeauthoryear{Butland et~al.}{2008}]{butland2008esga}
\begin{barticle}[author]
\bauthor{\bsnm{Butland},~\bfnm{Gareth}\binits{G.}},
  \bauthor{\bsnm{Babu},~\bfnm{Mohan}\binits{M.}},
  \bauthor{\bsnm{D{\'\i}az-Mej{\'\i}a},~\bfnm{J~Javier}\binits{J.~J.}},
  \bauthor{\bsnm{Bohdana},~\bfnm{Fedyshyn}\binits{F.}},
  \bauthor{\bsnm{Phanse},~\bfnm{Sadhna}\binits{S.}},
  \bauthor{\bsnm{Gold},~\bfnm{Barbara}\binits{B.}},
  \bauthor{\bsnm{Yang},~\bfnm{Wenhong}\binits{W.}},
  \bauthor{\bsnm{Li},~\bfnm{Joyce}\binits{J.}},
  \bauthor{\bsnm{Gagarinova},~\bfnm{Alla~G}\binits{A.~G.}},
  \bauthor{\bsnm{Pogoutse},~\bfnm{Oxana}\binits{O.}} \betal{et~al.}
(\byear{2008}).
\btitle{eSGA: E. coli synthetic genetic array analysis}.
\bjournal{Nature methods}
\bvolume{5}
\bpages{789--795}.
\end{barticle}
\endbibitem

\bibitem[\protect\citeauthoryear{Dowle and Srinivasan}{2018}]{datatable}
\begin{bmanual}[author]
\bauthor{\bsnm{Dowle},~\bfnm{Matt}\binits{M.}} \AND
  \bauthor{\bsnm{Srinivasan},~\bfnm{Arun}\binits{A.}}
(\byear{2018}).
\btitle{data.table: Extension of `data.frame`}
\bnote{R package version 1.11.8}.
\end{bmanual}
\endbibitem

\bibitem[\protect\citeauthoryear{Dudoit et~al.}{{2002}}]{dudoit2002microarray}
\begin{barticle}[author]
\bauthor{\bsnm{Dudoit},~\bfnm{S}\binits{S.}},
  \bauthor{\bsnm{Yang},~\bfnm{YH}\binits{Y.}},
  \bauthor{\bsnm{Callow},~\bfnm{MJ}\binits{M.}} \AND
  \bauthor{\bsnm{Speed},~\bfnm{TP}\binits{T.}}
(\byear{{2002}}).
\btitle{{Statistical methods for identifying differentially expressed genes in
  replicated cDNA microarray experiments}}.
\bjournal{{STATISTICA SINICA}}
\bvolume{{12}}
\bpages{{111-139}}.
\end{barticle}
\endbibitem

\bibitem[\protect\citeauthoryear{Efron et~al.}{2004}]{efron2004least}
\begin{barticle}[author]
\bauthor{\bsnm{Efron},~\bfnm{Bradley}\binits{B.}},
  \bauthor{\bsnm{Hastie},~\bfnm{Trevor}\binits{T.}},
  \bauthor{\bsnm{Johnstone},~\bfnm{Iain}\binits{I.}},
  \bauthor{\bsnm{Tibshirani},~\bfnm{Robert}\binits{R.}} \betal{et~al.}
(\byear{2004}).
\btitle{Least angle regression}.
\bjournal{The Annals of statistics}
\bvolume{32}
\bpages{407--499}.
\end{barticle}
\endbibitem

\bibitem[\protect\citeauthoryear{Ekstr{\o}m}{2018}]{MESS}
\begin{bmanual}[author]
\bauthor{\bsnm{Ekstr{\o}m},~\bfnm{Claus~Thorn}\binits{C.~T.}}
(\byear{2018}).
\btitle{MESS: Miscellaneous Esoteric Statistical Scripts}
\bnote{R package version 0.5.2}.
\end{bmanual}
\endbibitem

\bibitem[\protect\citeauthoryear{Florea and Vorobyov}{2017}]{florea2017robust}
\begin{binproceedings}[author]
\bauthor{\bsnm{Florea},~\bfnm{Mihai~I}\binits{M.~I.}} \AND
  \bauthor{\bsnm{Vorobyov},~\bfnm{Sergiy~A}\binits{S.~A.}}
(\byear{2017}).
\btitle{A robust FISTA-like algorithm}.
In \bbooktitle{2017 IEEE International Conference on Acoustics, Speech and
  Signal Processing (ICASSP)}
\bpages{4521--4525}.
\bpublisher{IEEE}.
\end{binproceedings}
\endbibitem

\bibitem[\protect\citeauthoryear{Friedman, Hastie and
  Tibshirani}{2010}]{friedman2010regularization}
\begin{barticle}[author]
\bauthor{\bsnm{Friedman},~\bfnm{Jerome}\binits{J.}},
  \bauthor{\bsnm{Hastie},~\bfnm{Trevor}\binits{T.}} \AND
  \bauthor{\bsnm{Tibshirani},~\bfnm{Rob}\binits{R.}}
(\byear{2010}).
\btitle{Regularization paths for generalized linear models via coordinate
  descent}.
\bjournal{Journal of statistical software}
\bvolume{33}
\bpages{1}.
\end{barticle}
\endbibitem

\bibitem[\protect\citeauthoryear{Fu}{1998}]{fu1998penalized}
\begin{barticle}[author]
\bauthor{\bsnm{Fu},~\bfnm{Wenjiang~J}\binits{W.~J.}}
(\byear{1998}).
\btitle{Penalized regressions: the bridge versus the lasso}.
\bjournal{Journal of computational and graphical statistics}
\bvolume{7}
\bpages{397--416}.
\end{barticle}
\endbibitem

\bibitem[\protect\citeauthoryear{Ghadimi et~al.}{2012}]{ghadimi2012optimal}
\begin{barticle}[author]
\bauthor{\bsnm{Ghadimi},~\bfnm{Euhanna}\binits{E.}},
  \bauthor{\bsnm{Teixeira},~\bfnm{Andr{\'e}}\binits{A.}},
  \bauthor{\bsnm{Shames},~\bfnm{Iman}\binits{I.}} \AND
  \bauthor{\bsnm{Johansson},~\bfnm{Mikael}\binits{M.}}
(\byear{2012}).
\btitle{On the optimal step-size selection for the alternating direction method
  of multipliers}.
\bjournal{IFAC Proceedings Volumes}
\bvolume{45}
\bpages{139--144}.
\end{barticle}
\endbibitem

\bibitem[\protect\citeauthoryear{Hobbs, Astarita and
  Storz}{2010}]{hobbs2010small}
\begin{barticle}[author]
\bauthor{\bsnm{Hobbs},~\bfnm{Errett~C}\binits{E.~C.}},
  \bauthor{\bsnm{Astarita},~\bfnm{Jillian~L}\binits{J.~L.}} \AND
  \bauthor{\bsnm{Storz},~\bfnm{Gisela}\binits{G.}}
(\byear{2010}).
\btitle{Small RNAs and small proteins involved in resistance to cell envelope
  stress and acid shock in Escherichia coli: analysis of a bar-coded mutant
  collection}.
\bjournal{Journal of bacteriology}
\bvolume{192}
\bpages{59--67}.
\end{barticle}
\endbibitem

\bibitem[\protect\citeauthoryear{Javanmard and
  Montanari}{2014}]{javanmard2014confidence}
\begin{barticle}[author]
\bauthor{\bsnm{Javanmard},~\bfnm{Adel}\binits{A.}} \AND
  \bauthor{\bsnm{Montanari},~\bfnm{Andrea}\binits{A.}}
(\byear{2014}).
\btitle{Confidence Intervals and Hypothesis Testing for High-dimensional
  Regression}.
\bjournal{J. Mach. Learn. Res.}
\bvolume{15}
\bpages{2869--2909}.
\end{barticle}
\endbibitem

\bibitem[\protect\citeauthoryear{Kim and Fessler}{2018}]{kim2018another}
\begin{barticle}[author]
\bauthor{\bsnm{Kim},~\bfnm{Donghwan}\binits{D.}} \AND
  \bauthor{\bsnm{Fessler},~\bfnm{Jeffrey~A}\binits{J.~A.}}
(\byear{2018}).
\btitle{Another look at the fast iterative shrinkage/thresholding algorithm
  (FISTA)}.
\bjournal{SIAM Journal on Optimization}
\bvolume{28}
\bpages{223--250}.
\end{barticle}
\endbibitem

\bibitem[\protect\citeauthoryear{Liang, Nichols and
  Sen}{2019}]{liang2019matrix}
\begin{barticle}[author]
\bauthor{\bsnm{Liang},~\bfnm{Jane~W}\binits{J.~W.}},
  \bauthor{\bsnm{Nichols},~\bfnm{Robert~J}\binits{R.~J.}} \AND
  \bauthor{\bsnm{Sen},~\bfnm{{\'S}aunak}\binits{{\'S}.}}
(\byear{2019}).
\btitle{Matrix linear models for high-throughput chemical genetic screens}.
\bjournal{Genetics}
\bvolume{212}
\bpages{1063--1073}.
\end{barticle}
\endbibitem

\bibitem[\protect\citeauthoryear{Liang and
  Sch{\"o}nlieb}{2018}]{liang2018improving}
\begin{barticle}[author]
\bauthor{\bsnm{Liang},~\bfnm{Jingwei}\binits{J.}} \AND
  \bauthor{\bsnm{Sch{\"o}nlieb},~\bfnm{Carola-Bibiane}\binits{C.-B.}}
(\byear{2018}).
\btitle{Improving FISTA: Faster, Smarter and Greedier}.
\bjournal{arXiv preprint arXiv:1811.01430}.
\end{barticle}
\endbibitem

\bibitem[\protect\citeauthoryear{Liang and Sen}{2021a}]{liang2021mlmfigs}
\begin{barticle}[author]
\bauthor{\bsnm{Liang},~\bfnm{Jane~W}\binits{J.~W.}} \AND
  \bauthor{\bsnm{Sen},~\bfnm{{\'S}aunak}\binits{{\'S}.}}
(\byear{2021}a).
\btitle{Supplemental figures for ``Sparse matrix linear models for structured
  high-throughput data.''}.
\end{barticle}
\endbibitem

\bibitem[\protect\citeauthoryear{Liang and Sen}{2021b}]{liang2021mlmpackage}
\begin{barticle}[author]
\bauthor{\bsnm{Liang},~\bfnm{Jane~W}\binits{J.~W.}} \AND
  \bauthor{\bsnm{Sen},~\bfnm{{\'S}aunak}\binits{{\'S}.}}
(\byear{2021}b).
\btitle{matrixLMnet.jl Julia package for ``Sparse matrix linear models for
  structured high-throughput data.''}.
\end{barticle}
\endbibitem

\bibitem[\protect\citeauthoryear{Liang and Sen}{2021c}]{liang2021mlmrepo}
\begin{barticle}[author]
\bauthor{\bsnm{Liang},~\bfnm{Jane~W}\binits{J.~W.}} \AND
  \bauthor{\bsnm{Sen},~\bfnm{{\'S}aunak}\binits{{\'S}.}}
(\byear{2021}c).
\btitle{Code to reproduce analysis for ``Sparse matrix linear models for
  structured high-throughput data.''}.
\end{barticle}
\endbibitem

\bibitem[\protect\citeauthoryear{Lovell et~al.}{2015}]{lovell2015exploiting}
\begin{barticle}[author]
\bauthor{\bsnm{Lovell},~\bfnm{John~T}\binits{J.~T.}},
  \bauthor{\bsnm{Mullen},~\bfnm{Jack~L}\binits{J.~L.}},
  \bauthor{\bsnm{Lowry},~\bfnm{David~B}\binits{D.~B.}},
  \bauthor{\bsnm{Awole},~\bfnm{Kedija}\binits{K.}},
  \bauthor{\bsnm{Richards},~\bfnm{James~H}\binits{J.~H.}},
  \bauthor{\bsnm{Sen},~\bfnm{Saunak}\binits{S.}},
  \bauthor{\bsnm{Verslues},~\bfnm{Paul~E}\binits{P.~E.}},
  \bauthor{\bsnm{Juenger},~\bfnm{Thomas~E}\binits{T.~E.}} \AND
  \bauthor{\bsnm{McKay},~\bfnm{John~K}\binits{J.~K.}}
(\byear{2015}).
\btitle{Exploiting differential gene expression and epistasis to discover
  candidate genes for drought-associated QTLs in Arabidopsis thaliana}.
\bjournal{The Plant Cell}
\bvolume{27}
\bpages{969--983}.
\end{barticle}
\endbibitem

\bibitem[\protect\citeauthoryear{Lowry et~al.}{2013}]{lowry2013expression}
\begin{barticle}[author]
\bauthor{\bsnm{Lowry},~\bfnm{David~B}\binits{D.~B.}},
  \bauthor{\bsnm{Logan},~\bfnm{Tierney~L}\binits{T.~L.}},
  \bauthor{\bsnm{Santuari},~\bfnm{Luca}\binits{L.}},
  \bauthor{\bsnm{Hardtke},~\bfnm{Christian~S}\binits{C.~S.}},
  \bauthor{\bsnm{Richards},~\bfnm{James~H}\binits{J.~H.}},
  \bauthor{\bsnm{DeRose-Wilson},~\bfnm{Leah~J}\binits{L.~J.}},
  \bauthor{\bsnm{McKay},~\bfnm{John~K}\binits{J.~K.}},
  \bauthor{\bsnm{Sen},~\bfnm{Saunak}\binits{S.}} \AND
  \bauthor{\bsnm{Juenger},~\bfnm{Thomas~E}\binits{T.~E.}}
(\byear{2013}).
\btitle{Expression quantitative trait locus mapping across water availability
  environments reveals contrasting associations with genomic features in
  Arabidopsis}.
\bjournal{The Plant Cell}
\bvolume{25}
\bpages{3266--3279}.
\end{barticle}
\endbibitem

\bibitem[\protect\citeauthoryear{Nesterov}{1983}]{nesterov1983method}
\begin{binproceedings}[author]
\bauthor{\bsnm{Nesterov},~\bfnm{Yurii}\binits{Y.}}
(\byear{1983}).
\btitle{A method of solving a convex programming problem with convergence rate
  {O} (1/k2)}.
In \bbooktitle{Soviet Mathematics Doklady}
\bvolume{27}
\bpages{372--376}.
\end{binproceedings}
\endbibitem

\bibitem[\protect\citeauthoryear{Nichols et~al.}{2011}]{nichols2011phenotypic}
\begin{barticle}[author]
\bauthor{\bsnm{Nichols},~\bfnm{Robert~J.}\binits{R.~J.}},
  \bauthor{\bsnm{Sen},~\bfnm{\'{S}aunak}\binits{S.}},
  \bauthor{\bsnm{Choo},~\bfnm{Yoe~Jin}\binits{Y.~J.}},
  \bauthor{\bsnm{Beltrao},~\bfnm{Pedro}\binits{P.}},
  \bauthor{\bsnm{Zietek},~\bfnm{Matylda}\binits{M.}},
  \bauthor{\bsnm{Chaba},~\bfnm{Rachna}\binits{R.}},
  \bauthor{\bsnm{Lee},~\bfnm{Sueyoung}\binits{S.}},
  \bauthor{\bsnm{Kazmierczak},~\bfnm{Krystyna~M.}\binits{K.~M.}},
  \bauthor{\bsnm{Lee},~\bfnm{Karis~J.}\binits{K.~J.}},
  \bauthor{\bsnm{Wong},~\bfnm{Angela}\binits{A.}} \betal{et~al.}
(\byear{2011}).
\btitle{Phenotypic landscape of a bacterial cell}.
\bjournal{Cell}
\bvolume{144}
\bpages{143--156}.
\end{barticle}
\endbibitem

\bibitem[\protect\citeauthoryear{Ochs and Pock}{2017}]{ochs2017adaptive}
\begin{barticle}[author]
\bauthor{\bsnm{Ochs},~\bfnm{Peter}\binits{P.}} \AND
  \bauthor{\bsnm{Pock},~\bfnm{Thomas}\binits{T.}}
(\byear{2017}).
\btitle{Adaptive FISTA for Non-convex Optimization}.
\bjournal{arXiv preprint arXiv:1711.04343}.
\end{barticle}
\endbibitem

\bibitem[\protect\citeauthoryear{Parikh and Boyd}{2014}]{parikh2014proximal}
\begin{barticle}[author]
\bauthor{\bsnm{Parikh},~\bfnm{N}\binits{N.}} \AND
  \bauthor{\bsnm{Boyd},~\bfnm{S}\binits{S.}}
(\byear{2014}).
\btitle{Proximal algorithms}.
\bjournal{Foundations and Trends in Optimization}
\bvolume{1}
\bpages{123--231}.
\end{barticle}
\endbibitem

\bibitem[\protect\citeauthoryear{Ramsay and
  Silverman}{2005}]{ramsay2005functional}
\begin{bbook}[author]
\bauthor{\bsnm{Ramsay},~\bfnm{J.~O.}\binits{J.~O.}} \AND
  \bauthor{\bsnm{Silverman},~\bfnm{B.~W.}\binits{B.~W.}}
(\byear{2005}).
\btitle{Functional data analysis},
\bedition{2nd ed} ed.
\bseries{Springer series in statistics}.
\bpublisher{Springer}, \baddress{New York}.
\end{bbook}
\endbibitem

\bibitem[\protect\citeauthoryear{Reid, Tibshirani and
  Friedman}{2016}]{reid2016lasso}
\begin{barticle}[author]
\bauthor{\bsnm{Reid},~\bfnm{Stephen}\binits{S.}},
  \bauthor{\bsnm{Tibshirani},~\bfnm{Robert}\binits{R.}} \AND
  \bauthor{\bsnm{Friedman},~\bfnm{Jerome}\binits{J.}}
(\byear{2016}).
\btitle{A study of error variance estimation in {Lasso} regression}.
\bjournal{Statistica Sinica}.
\bdoi{10.5705/ss.2014.042}
\end{barticle}
\endbibitem

\bibitem[\protect\citeauthoryear{Ritchie et~al.}{2015}]{ritchie2015limma}
\begin{barticle}[author]
\bauthor{\bsnm{Ritchie},~\bfnm{Matthew~E.}\binits{M.~E.}},
  \bauthor{\bsnm{Phipson},~\bfnm{Belinda}\binits{B.}},
  \bauthor{\bsnm{Wu},~\bfnm{Di}\binits{D.}},
  \bauthor{\bsnm{Hu},~\bfnm{Yifang}\binits{Y.}},
  \bauthor{\bsnm{Law},~\bfnm{Charity~W.}\binits{C.~W.}},
  \bauthor{\bsnm{Shi},~\bfnm{Wei}\binits{W.}} \AND
  \bauthor{\bsnm{Smyth},~\bfnm{Gordon~K.}\binits{G.~K.}}
(\byear{2015}).
\btitle{limma powers differential expression analyses for {RNA}-sequencing and
  microarray studies}.
\bjournal{Nucleic Acids Research}
\bvolume{43}
\bpages{e47--e47}.
\bdoi{10.1093/nar/gkv007}
\end{barticle}
\endbibitem

\bibitem[\protect\citeauthoryear{Schmidt, Fung and
  Rosales}{2009}]{schmidt2009optimization}
\begin{barticle}[author]
\bauthor{\bsnm{Schmidt},~\bfnm{Mark}\binits{M.}},
  \bauthor{\bsnm{Fung},~\bfnm{Glenn}\binits{G.}} \AND
  \bauthor{\bsnm{Rosales},~\bfnm{Romer}\binits{R.}}
(\byear{2009}).
\btitle{Optimization methods for l1-regularization}.
\bjournal{University of British Columbia, Technical Report TR-2009}
\bvolume{19}.
\end{barticle}
\endbibitem

\bibitem[\protect\citeauthoryear{Su, Boyd and
  Cand\`es}{2016}]{su2016differential}
\begin{barticle}[author]
\bauthor{\bsnm{Su},~\bfnm{Weijie}\binits{W.}},
  \bauthor{\bsnm{Boyd},~\bfnm{Stephen}\binits{S.}} \AND
  \bauthor{\bsnm{Cand\`es},~\bfnm{Emmanuel~J}\binits{E.~J.}}
(\byear{2016}).
\btitle{A Differential Equation for Modeling {N}esterov's Accelerated Gradient
  Method: {T}heory and Insights}.
\bjournal{Journal of Machine Learning Research}
\bvolume{17}
\bpages{1--43}.
\end{barticle}
\endbibitem

\bibitem[\protect\citeauthoryear{Subramanian
  et~al.}{2005}]{subramanian2005gsea}
\begin{barticle}[author]
\bauthor{\bsnm{Subramanian},~\bfnm{Aravind}\binits{A.}},
  \bauthor{\bsnm{Tamayo},~\bfnm{Pablo}\binits{P.}},
  \bauthor{\bsnm{Mootha},~\bfnm{Vamsi~K.}\binits{V.~K.}},
  \bauthor{\bsnm{Mukherjee},~\bfnm{Sayan}\binits{S.}},
  \bauthor{\bsnm{Ebert},~\bfnm{Benjamin~L.}\binits{B.~L.}},
  \bauthor{\bsnm{Gillette},~\bfnm{Michael~A.}\binits{M.~A.}},
  \bauthor{\bsnm{Paulovich},~\bfnm{Amanda}\binits{A.}},
  \bauthor{\bsnm{Pomeroy},~\bfnm{Scott~L.}\binits{S.~L.}},
  \bauthor{\bsnm{Golub},~\bfnm{Todd~R.}\binits{T.~R.}},
  \bauthor{\bsnm{Lander},~\bfnm{Eric~S.}\binits{E.~S.}} \AND
  \bauthor{\bsnm{Mesirov},~\bfnm{Jill~P.}\binits{J.~P.}}
(\byear{2005}).
\btitle{Gene set enrichment analysis: {A} knowledge-based approach for
  interpreting genome-wide expression profiles}.
\bjournal{Proceedings of the National Academy of Sciences of the United States
  of America}
\bvolume{102}
\bpages{15545--15550}.
\bdoi{10.1073/pnas.0506580102}
\end{barticle}
\endbibitem

\bibitem[\protect\citeauthoryear{Tan et~al.}{2014}]{tan2014hubs}
\begin{barticle}[author]
\bauthor{\bsnm{Tan},~\bfnm{Kean~Ming}\binits{K.~M.}},
  \bauthor{\bsnm{London},~\bfnm{Palma}\binits{P.}},
  \bauthor{\bsnm{Mohan},~\bfnm{Karthik}\binits{K.}},
  \bauthor{\bsnm{Lee},~\bfnm{Su-In}\binits{S.-I.}},
  \bauthor{\bsnm{Fazel},~\bfnm{Maryam}\binits{M.}} \AND
  \bauthor{\bsnm{Witten},~\bfnm{Daniela}\binits{D.}}
(\byear{2014}).
\btitle{Learning {Graphical} {Models} {With} {Hubs}}.
\bjournal{Journal of machine learning research: JMLR}
\bvolume{15}
\bpages{3297--3331}.
\end{barticle}
\endbibitem

\bibitem[\protect\citeauthoryear{{R Core Team}}{2018}]{R}
\begin{bmanual}[author]
\bauthor{\bsnm{{R Core Team}}}
(\byear{2018}).
\btitle{R: A Language and Environment for Statistical Computing}
\bpublisher{R Foundation for Statistical Computing},
\baddress{Vienna, Austria}.
\end{bmanual}
\endbibitem

\bibitem[\protect\citeauthoryear{Team et~al.}{2017}]{mutoss}
\begin{bmanual}[author]
\bauthor{\bsnm{Team},~\bfnm{MuToss~Coding}\binits{M.~C.}},
  \bauthor{\bsnm{Blanchard},~\bfnm{Gilles}\binits{G.}},
  \bauthor{\bsnm{Dickhaus},~\bfnm{Thorsten}\binits{T.}},
  \bauthor{\bsnm{Hack},~\bfnm{Niklas}\binits{N.}},
  \bauthor{\bsnm{Konietschke},~\bfnm{Frank}\binits{F.}},
  \bauthor{\bsnm{Rohmeyer},~\bfnm{Kornelius}\binits{K.}},
  \bauthor{\bsnm{Rosenblatt},~\bfnm{Jonathan}\binits{J.}},
  \bauthor{\bsnm{Scheer},~\bfnm{Marsel}\binits{M.}} \AND
  \bauthor{\bsnm{Werft},~\bfnm{Wiebke}\binits{W.}}
(\byear{2017}).
\btitle{mutoss: Unified Multiple Testing Procedures}
\bnote{R package version 0.1-12}.
\end{bmanual}
\endbibitem

\bibitem[\protect\citeauthoryear{Tibshirani}{1996}]{tibshirani1996regression}
\begin{barticle}[author]
\bauthor{\bsnm{Tibshirani},~\bfnm{Robert}\binits{R.}}
(\byear{1996}).
\btitle{Regression shrinkage and selection via the lasso}.
\bjournal{Journal of the Royal Statistical Society: Series B (Methodological)}
\bvolume{58}
\bpages{267--288}.
\end{barticle}
\endbibitem

\bibitem[\protect\citeauthoryear{Woodruff, Zota and
  Schwartz}{2011}]{woodruff2011environmental}
\begin{barticle}[author]
\bauthor{\bsnm{Woodruff},~\bfnm{Tracey~J.}\binits{T.~J.}},
  \bauthor{\bsnm{Zota},~\bfnm{Ami~R.}\binits{A.~R.}} \AND
  \bauthor{\bsnm{Schwartz},~\bfnm{Jackie~M.}\binits{J.~M.}}
(\byear{2011}).
\btitle{Environmental chemicals in pregnant women in the {U}nited {S}tates:
  {NHANES} 2003-2004}.
\bjournal{Environmental health perspectives}
\bvolume{119}
\bpages{878}.
\end{barticle}
\endbibitem

\bibitem[\protect\citeauthoryear{Wu and Lange}{2008}]{wu2008coordinate}
\begin{barticle}[author]
\bauthor{\bsnm{Wu},~\bfnm{Tong~Tong}\binits{T.~T.}} \AND
  \bauthor{\bsnm{Lange},~\bfnm{Kenneth}\binits{K.}}
(\byear{2008}).
\btitle{Coordinate descent algorithms for lasso penalized regression}.
\bjournal{The Annals of Applied Statistics}
\bpages{224--244}.
\end{barticle}
\endbibitem

\bibitem[\protect\citeauthoryear{Xiong et~al.}{2011}]{xiong2011flexible}
\begin{barticle}[author]
\bauthor{\bsnm{Xiong},~\bfnm{Hao}\binits{H.}},
  \bauthor{\bsnm{Goulding},~\bfnm{Evan~H}\binits{E.~H.}},
  \bauthor{\bsnm{Carlson},~\bfnm{Elaine~J}\binits{E.~J.}},
  \bauthor{\bsnm{Tecott},~\bfnm{Laurence~H}\binits{L.~H.}},
  \bauthor{\bsnm{McCulloch},~\bfnm{Charles~E}\binits{C.~E.}} \AND
  \bauthor{\bsnm{Sen},~\bfnm{{\'S}aunak}\binits{{\'S}.}}
(\byear{2011}).
\btitle{A flexible estimating equations approach for mapping function-valued
  traits}.
\bjournal{Genetics}
\bvolume{189}
\bpages{305--316}.
\end{barticle}
\endbibitem

\bibitem[\protect\citeauthoryear{Zou and Hastie}{2005}]{zou2005regularization}
\begin{barticle}[author]
\bauthor{\bsnm{Zou},~\bfnm{Hui}\binits{H.}} \AND
  \bauthor{\bsnm{Hastie},~\bfnm{Trevor}\binits{T.}}
(\byear{2005}).
\btitle{Regularization and variable selection via the elastic net}.
\bjournal{Journal of the Royal Statistical Society: Series B (Statistical
  Methodology)}
\bvolume{67}
\bpages{301--320}.
\end{barticle}
\endbibitem

\end{thebibliography}

\end{document}